\documentclass{ieeeaccess}

\usepackage{fancyhdr}

\usepackage[OT1]{fontenc}
\usepackage[cmex10]{amsmath}
\usepackage{amssymb}
\usepackage{bm}
\usepackage{graphicx}
\usepackage{subfigure}
\usepackage{tabularx}
\usepackage{arydshln}
\usepackage{mathtools}
\usepackage{multirow}
\usepackage{algorithm,algorithmic}
\usepackage{url}
\usepackage{color}
\usepackage{comment}
\usepackage{amsthm}
\usepackage{braket}
\usepackage{physics}
\usepackage{lipsum}
\usepackage{cite}
\usepackage{multicol}
\usepackage{float}
\usepackage{caption}
\usepackage[absolute,overlay]{textpos}

\graphicspath{{./figure_final/}}

\RequirePackage[normalem]{ulem} 
\RequirePackage{color}\definecolor{RED}{rgb}{1,0,0}\definecolor{BLUE}{rgb}{0,0,1} 

\providecommand{\DIFdelFL}[1]{} 

\newcommand{\argmin}{\mathop{\mathrm{arg~min}}\limits}

\newcommand{\inbb}[2]{\in \mathbb{#1}^{#2}}

\def\BibTeX{{\rm B\kern-.05em{\sc i\kern-.025em b}\kern-.08em
    T\kern-.1667em\lower.7ex\hbox{E}\kern-.125emX}}

\begin{document}
\doi{10.1109/TQE.2025.3595910}

\title{Grover Adaptive Search with Spin Variables}
\author{
\uppercase{Shintaro~Fujiwara}\authorrefmark{1},
\IEEEmembership{Graduate Student~Member,~IEEE},
\uppercase{Naoki~Ishikawa}\authorrefmark{1}, \IEEEmembership{Senior~Member,~IEEE}
}
\address[1]{Faculty of Engineering, Yokohama National University, Yokohama, 240-8501 Kanagawa, Japan}
\tfootnote{This work was supported by JST CRONOS Japan Grant Number JPMJCS24N1.}

\markboth
{Author \headeretal: Preparation of Papers for IEEE Transactions on Quantum Engineering}
{Author \headeretal: Preparation of Papers for IEEE Transactions on Quantum Engineering}

\corresp{Corresponding author: Shintaro Fujiwara (email: fujiwara-shintaro-by@ynu.jp).}

\begin{abstract}
This paper presents a novel approach to Grover adaptive search (GAS) for a combinatorial optimization problem whose objective function involves spin variables.
While the GAS algorithm with a conventional design of a quantum dictionary subroutine handles a problem associated with an objective function with binary variables $\{0,1\}$, we reformulate the problem using spin variables $\{+1,-1\}$ to simplify the algorithm. Specifically, we introduce a novel quantum dictionary subroutine that is designed for this spin-based formulation. 
A key benefit of this approach is the substantial reduction in the number of CNOT gates required to construct the quantum circuit. We theoretically demonstrate that, for certain problems, our proposed approach can reduce the gate complexity from an exponential order to a polynomial order, compared to the conventional binary-based approach. This improvement has the potential to enhance the scalability and efficiency of GAS, particularly in larger quantum computations.
\end{abstract}


\maketitle

\fancypagestyle{firstpage}{%
  \fancyhf{}
  \fancyfoot[C]{\scriptsize
  \textcopyright~2025 IEEE. Personal use of this material is permitted.
  Permission from IEEE must be obtained for all other uses, in any current or future
  media, including reprinting/republishing this material for advertising or promotional
  purposes, creating new collective works, for resale or redistribution to servers or
  lists, or reuse of any copyrighted component of this work in other works. 
  DOI: 10.1109/TQE.2025.3595910}
}
\thispagestyle{firstpage}

\TPshowboxesfalse
\begin{textblock*}{\textwidth}(45pt,10pt)
\footnotesize
\centering
Accepted for publication in IEEE Transactions on Quantum Engineering. This is the author's version which has not been fully edited and content may be different from the final publication. Citation information: DOI: 10.1109/TQE.2025.3595910
\end{textblock*}

\section{Introduction}
\IEEEPARstart{C}{onstrained} polynomial binary optimization (CPBO) problems have been addressed in a wide range of fields such as finance \cite{egger2020quantum}, healthcare \cite{iasemidis2001quadratic}, wireless communications \cite{sano2023qubit}, etc. 
In classical computing, CPBO problems are typically solved using state-of-the-art mathematical programming solvers \cite{ibm2022user, gurobioptimizationllc2023gurobi}
 as well as efficient tools using the semidefinite relaxation \cite{luo2010semidefinite}, metaheuristics, approximation algorithms, and machine learning \cite{ning2019optimization}.
In the context of quantum computing, quantum annealing (QA) \cite{kadowaki1998quantum} and the quantum approximate optimization algorithm (QAOA) \cite{farhi2014quantum} are popular approaches. Both rely on the quantum adiabatic theorem, with QA being capable of solving quadratic unconstrained binary optimization (QUBO) problems, while QAOA extends this capability to also solve higher-order cases, i.e., higher-order unconstrained binary optimization (HUBO) problems \cite{campbell2021qaoa}. 
Despite their high feasibility for extensive practical implementations on real quantum devices \cite{johnson2011quantum,andrew2019validating}, 
it has been demonstrated that quantum advantages are unlikely for classical optimization due to their high noise rates \cite{stilckfranca2021limitations}.

Assuming the future realization of fault-tolerant quantum computing (FTQC), which can overcome the negative effects of gate noise, a particularly promising approach is the Grover adaptive search (GAS) \cite{gilliam2021grover,gilliam2020optimizing}.
The GAS algorithm minimizes a given polynomial binary function using Grover's search algorithm \cite{grover1996fast}, which can find an exact solution in a search space of size $2^n$ with query complexity $O(\sqrt{2^n})$~\footnote{$n$ is a positive integer representing the number of qubits, and the search space contains $2^n$ items.}, providing a quadratic speedup for solving QUBO and HUBO problems.
Gilliam's construction method \cite{gilliam2021grover} originally assumes polynomial coefficients to be integers. 
However, the study \cite{norimoto2023quantum} has extended it to handle polynomials with real coefficients. 
The quantum circuit for the GAS algorithm consists of $n+m$ qubits, where $n$ is the number of binary variables and $m$ is the number of qubits required to encode objective function values using the proposed quantum dictionary \cite{gilliam2021foundational}.
To simplify the circuit, if the objective function is in a QUBO form, Nagy et al. proposed a technique that replaces the multi-control phase gate in the circuit with a combination of CNOT and $\mathbf{R}_{\mathrm{z}}$ gates, thereby improving the implementation efficiency \cite{nagy2024fixedpoint}.

In formulating nondeterministic polynomial-time (NP) problems, one-hot encoding that relies on one-hot vectors is often used \cite{lucas2014ising}. In contrast, binary encoding, which converts a one-hot vector into a binary sequence, is known to reduce the number of qubits required \cite{fuchs2021efficient,glos2022spaceefficient} as well as the total number of gates by canceling some of them \cite{sano2024accelerating}. Additionally, by optimizing algorithm parameters, query complexity can be further reduced \cite{giuffrida2022engineering,ominato2024grover}, although it has been proven that it is impossible to be reduced below $o(\sqrt{2^n})$ \cite{bennett1997strengths}\footnote{Let $f(x)$ and $g(x)$ be functions defined near the real value $a$. If there exists a constant $C$ such that $\left|\frac{f(x)}{g(x)}\right| \leq C$ as $x$ approaches $a$, then $f(x)$ is described using big-$O$ notation, i.e., $f(x) = O(g(x))$, meaning that $f(x)$ converges at the same rate or slower than $g(x)$. Conversely, if $\lim_{x \to a} \left|\frac{f(x)}{g(x)}\right| = 0$, then $f(x)$ is described using little-$o$ notation, namely $f(x) = o(g(x))$, which means that $f(x)$ converges slower than $g(x)$ \cite{Jones1997Computability}.}.
Since the proposal of the GAS algorithm, its applications have been studied in various fields, 
including
traveling salesman problem (TSP) \cite{zhu2022realizable,ominato2024grover},
industrial shift scheduling \cite{Krol2024QISS}, detection problems in wireless communications \cite{norimoto2023quantum, norimoto2024quantum}, channel allocation problems \cite{sano2023qubit}, 
the construction of binary constant weight codes \cite{yukiyoshi2022quantum},
and the dispersion problem \cite{yukiyoshi2024quantum}.

In the quantum circuit for the GAS algorithm, all variables are represented in binary as $\{0, 1\}$. Here, using spin variables $\{+1, -1\}$ instead of binary variables $\{0, 1\}$ may result in a simpler objective function in some cases. For instance, when considering the decoding problem of error-correcting codes, the objective function can be expressed using the Ising model, where spin variables $\{+1, -1\}$ are used. Let us assume that the variables $+1$ and $-1$ correspond to $0$ and $1$ in the Galois field $\mathbb{F}_2$, respectively. In this case, we notice that the XOR operation for binary variables directly corresponds to the multiplication of these spin variables. This allows for a more concise representation of the XOR operation and reduces the number of terms in the objective function.
Specifically, when we try to solve the problem using binary variables, a spin variable $s \in \{+1, -1\}$ has to be replaced by $s = 1 - 2x$ with $x \in \{0, 1\}$, and the multiplication of the polynomials in the form of $1-2x$ increases the number of terms in the objective function exponentially. 
A typical issue can be found in the application of the GAS algorithm for the maximum-likelihood~(ML) detection of wireless systems \cite{norimoto2023quantum}. 
To represent Gray-coded QAM symbols, higher-order polynomials are specified in the 5G standard \cite{3gpp2018ts}, but the Gray codes are basically generated using XOR operations, making the objective function more concise with spin variables. Applying the GAS algorithm with the conventional circuit design to problems involving XOR operations causes  the number of terms in the objective function to increase significantly, which can disrupt the feasibility of the GAS algorithm.

Additionally, the GAS quantum circuit has the potential to further enhance efficiency.
For an $n$-th order term, the quantum dictionary subroutine requires $m$ phase gates controlled by $n$ qubits. Decomposing each of these multi-control phase gates into elementary gates without an ancilla qubit requires $O(n^2)$ CNOT gates even with a state-of-the-art method \cite{Silva2023LinearDecomposition}, as the phase gate is in $\mathcal{U}(2)$.

Against this background, this paper proposes a method to construct a GAS quantum circuit using spin variables $s \in \{+1, -1\}$ instead of binary variables $x \in \{0, 1\}$. We demonstrate that it is possible to compute objective function values with spin variables by replacing multi-control phase gates with a combination of rotation-Z gates and CNOT gates. This approach is particularly effective on superconducting hardware, where the rotation-Z gate is often implemented as a ``virtual Z gate'', a software-based gate operation with effectively zero duration and gate error\cite{McKay2017Efficient}. 
Furthermore, this modification reduces the total number of CNOT gates, potentially improving the feasibility of the GAS algorithm.
The contributions of this paper are summarized as follows.
\begin{enumerate}
\item We propose a novel design of the quantum dictionary circuit that encodes objective function values using spin variables. This circuit is composed of CNOT and rotation-Z gates, enabling efficient implementation, whereas the conventional quantum dictionary relies on multi-controlled phase gates which require a larger number of elementary gates to be decomposed.

\item We demonstrate that the objective functions for solving an ML detection problem and a syndrome decoding problem can be significantly simplified by using the proposed circuit design of the quantum dictionary instead of the conventional design. 
Specifically, the number of terms in these objective functions can be reduced from an exponential order to a polynomial order, which can increase the feasibility of the GAS algorithm.
\end{enumerate}

It is important to note that a new design of the quantum dictionary, based solely on CNOT and $\mathbf{R}_\mathrm{z}$ gates, for encoding the objective function values with binary variables in QUBO form including XOR operations was independently proposed in \cite{nagy2024fixedpoint}, along with the introduction of fixed-point Grover search into the GAS algorithm. 
In contrast, the quantum dictionary design proposed in this paper supports objective functions in HUBO form. 
Furthermore, in this paper, we discuss the advantages of the GAS algorithm with the spin-based circuit design used for practical use cases, which have not been explored in the literature.

The rest of this paper is organized as follows.
In Section~\ref{sec:conv}, we provide the preliminaries of the GAS algorithm and the construction of the quantum circuit for the quantum dictionary.
In Section~\ref{sec:prop}, we propose the method to construct the novel quantum circuit of the quantum dictionary corresponding to spin variables.
In Section~\ref{sec:gate_reduction}, we analyze the effects of gate reduction by our proposed method on certain use cases.
In Section~\ref{sec:performance}, we evaluate the proposed method by comparing the number of CNOT gates in the quantum circuit.
Finally, we conclude the paper in Section~\ref{sec:conc}.
The important notations used in this paper are summarized in the NOMENCLATURE.

\begin{table}[tbp]
	\centering
	\caption*{NOMENCLATURE\label{table:sym}}
	\begin{tabular}{lll}
	    $\mathbb{R}$ & & Real numbers \\
	    $\mathbb{C}$ & & Complex numbers \\
	    $\mathbb{Z}$ & & Integers \\
        $\mathbb{F}_2$ & & Binary numbers on finite binary field \\
        $\mathrm{j}$ & $\inbb{C}{}$ & Imaginary number  \\
	    $N_{\mathrm{t}}$ & $\inbb{Z}{}$ & Number of transmit antennas\\
		$N_{\mathrm{r}}$ & $\inbb{Z}{}$ & Number of receive antennas\\
		$\sigma^2$ & $\inbb {R}{}$ & Noise variance\\
		$\gamma$ & $\inbb {R}{}$ & Signal-to-noise ratio \\
		$E(\cdot)$ & $\inbb{Z}{}$ & Objective function \\
		$n$ & $\inbb{Z}{}$ & Number of binary variables\\
		$m$ & $\inbb{Z}{}$ & Number of qubits required to encode $E(\cdot)$ \\
		$c$ & $\inbb{Z}{}$ & Index of GAS iterations \\
            $d$ & $\inbb{Z}{}$ & Max Grover rotation counts \\
		$y, y_c$ & $\inbb{Z}{}$ & Threshold that is adaptively updated by GAS\\
		$L, L_c$ & $\inbb{Z}{}$ & Number of Grover operators\\
		$\mathbf{b}, \mathbf{b}_i$ & $\inbb {B}{n}$ & Binary variables\\
		$\mathbf{t}$ & $\inbb {C}{N_{\mathrm{t}}\times 1}$ & Transmitted symbol vector\\
		$\mathbf{r}$ & $\inbb {C}{N_{\mathrm{r}}\times 1}$ & Received symbol vector \\
		$\mathbf{H}_{\mathrm{c}}$ & $\inbb {C}{N_{\mathrm{r}}\times N_{\mathrm{t}}}$ & Channel coefficients \\
		$\mathbf{v}$ & $\inbb {C}{N_{\mathrm{r}}\times 1}$ & Additive white Gaussian noise vector\\
	\end{tabular}
\end{table}

\section{Conventional Grover Adaptive Search}
\label{sec:conv}
In this section, we review a construction method for the quantum dictionary used in GAS \cite{gilliam2021grover}.
Then, we explain the overall algorithm of GAS \cite{gilliam2021grover} for solving a combinatorial binary optimization problem.

\subsection{Overview of GAS Algorithm}

Suppose that a HUBO function with $n$ binary variables is given in
\begin{equation}\label{eq:hubo_objfun}
E(\mathbf{x}) = \sum_{\mathcal{I}\in\mathcal{T}} a_{\mathcal{I}} \prod_{i\in\mathcal{I}} x_i.
\end{equation}
Here, for $\mathbf{x} = [x_0, \cdots, x_{n-1}]^{\mathrm{T}} \in \{0, 1\}^n$, the set $\mathcal{I}$ denotes the collection of indices of the variables included in a term of the function, $\mathcal{T}$ denotes the set of all such index sets corresponding to every term, and $a_{\mathcal{I}}$ denotes the nonzero real coefficient associated with the term corresponding to $\mathcal{I}$.
Therefore, $|\mathcal{T}|$ corresponds to the number of non-zero terms in $E(\mathbf{x})$.
Now, with GAS, let us solve an unconstrained binary optimization problem of
\begin{align}
    \begin{split}
        \min& \quad E(\mathbf{x})\\
        \textrm{s.t.}& \quad x_i \in \{0, 1\} ~ (i = 0, \cdots, n-1).
    \end{split}
\end{align}

The GAS algorithm depends on the state preparation operator $\mathbf{A}_y$ corresponding to the objective function $E(\mathbf{x})-y$, where $y$ is a tentative threshold value, and the Grover operator $\mathbf{G} = \mathbf{A}_y \mathbf{D} \mathbf{A}_y^{\mathrm{H}} \mathbf{O}$ that amplifies the states of interest.
Here, the Grover diffusion operator $\mathbf{D}$ $\in \mathbb{R}^{n \times n}$  is defined as \cite{grover1996fast}
\begin{align}
    \mathbf{D}_{p,q} =
    \begin{cases}
        \frac{2}{N} & (p \ne q)\\
        -1 + \frac{2}{N} & (p = q)
    \end{cases}, (p,q = 0,1,\cdots, n-1),
    \label{eq:D-def}
\end{align}
where $N=2^n$.
The oracle $\mathbf{O}$ flips the phase of the states of interest, whose most significant qubit in the value part is $\ket{1}$, i.e., the states that satisfy $E(\mathbf{x}) - y < 0$, which can be easily implemented by a single $\mathbf{Z}$ gate.
In this paper, we set the common ratio for increasing the Grover rotation count  to $\lambda = 8/7$, a constant real value within the range $1 < \lambda < 4/3$.

The GAS algorithm consists of the following four steps \cite{gilliam2021grover, norimoto2023quantum}.
\begin{enumerate}
    \item \textbf{Initialization}:
    A random solution $\mathbf{x}_0 \in \{0, 1\}^n$ is selected, and an initial threshold $y_0 = E(\mathbf{x}_0)$ is calculated classically.
    An iteration is set to $c=0$, and the maximum number of Grover rotation count $L_c$ is set to $d=1$.

    \item \textbf{Amplification}:
    The Grover rotation count $L_c$ is randomly selected from the set  $\{0, 1, \cdots, \lceil d-1 \rceil\}$, and the quantum state of $\mathbf{G}^{L_c} \mathbf{A}_{y_c} \Ket{0}_{n+m}$ is measured to obtain a solution $\mathbf{x}$. Its objective function value $y = E(\mathbf{x})$ is calculated in the classical domain.

    \item \textbf{Update}:
    If $y < y_c$, set $y_{c+1} = y$, $\mathbf{x}_{c+1} = \mathbf{x}$, and $d = 1$. Otherwise, set $y_{c+1} = y_c$, $\mathbf{x}_{c+1} = \mathbf{x}_c$, and $d = \min{\{\lambda d, \sqrt{2^n}\}}$. Then, update $c = c+1$.

    \item \textbf{Repetition}:
    The steps 2) and 3) are repeated until a termination condition is met.
\end{enumerate}
If the coefficients of the objective function are real values, the measurement probabilities of binary strings follow the Fejér distribution \cite{gilliam2021grover}, which indicates that multiple integers might become smaller than the true real-valued minimum of the objective function. 
As a result, the GAS algorithm may fail to reach an optimal solution because an incorrect integer value could be selected as the minimum and set as the new threshold $y_c$. 
However, this issue can be avoided by evaluating the objective function values in the classical domain in the Step 2), which can correctly evaluate the value of the objective function with real coefficients \cite{norimoto2023quantum}. 
Because we will focus on practical use cases where the objective function includes real-valued coefficients, as will be discussed in Sections~\ref{sec:gate_reduction} and \ref{sec:performance}, we adopt the algorithm \cite{norimoto2023quantum} in this paper.

\subsection{Quantum Dictionary for Calculating Objective Function Values}
\label{subsec:qd}

In a quantum circuit of GAS, a value of the objective function $E(\mathbf{x})$ can be encoded by the quantum dictionary proposed in \cite{gilliam2021foundational} .
The value of the objective function is represented by $m$ qubits, where $m$ satisfies
\begin{align}
    -2^{m - 1} \leq \min_{\mathbf{x}} E(\mathbf{x}) \leq \max_{\mathbf{x}} E(\mathbf{x}) < 2^{m - 1}.
\end{align}
Later, an operator $\mathbf{A}$ that encodes all possible values of $E(\mathbf{x})$ is constructed, i.e.,
\begin{align}\label{eq:A}
    \mathbf{A} \Ket{0}_{n+m} = \frac{1}{\sqrt{2^{n}}}\sum_{\mathbf{x} \in \{0, 1\}^{n}} \Ket{\mathbf{x}}_{n}\Ket{E(\mathbf{x})}_{m}.
\end{align}

Among the $n+m$ qubits, the first $n$ qubits are referred to as \textit{key part}, and the remaining $m$ qubits are referred to as \textit{value part}.
When $i$-th qubit in the key part is measured in the computational basis, an outcome of $0$ corresponds to $x_i=0$, and an outcome of $1$ corresponds to $x_i=1$.
Hadamard gates are applied to an all-zero quantum state, $\ket{0}^{\otimes(n+m)} = \ket{0}_{n+m}$, and its transition is
\begin{align}
\begin{split}
    \Ket{0}_{n+m}
    \xrightarrow{\mathbf{H}^{\otimes (n+m)}}
    &\frac{1}{\sqrt{2^{n+m}}} (\Ket{0}+\Ket{1})_{n+m}\\
    = &\frac{1}{\sqrt{2^{n + m}}} \sum_{\mathbf{x} \in \{0, 1\}^{n}} \Ket{\mathbf{x}}_{n}(\Ket{0}+\Ket{1})_{m}.
\end{split}
\label{eq:AyH}
\end{align}
Here, the tensor product of Hadamard gates is defined as
\begin{align}
    \mathbf{H}^{\otimes (n+m)} = \underbrace{\mathbf{H} \otimes \cdots \otimes \mathbf{H}}_{n+m} ~~ \text{and} ~~ \mathbf{H} = \frac{1}{\sqrt{2}}\mqty[1 & 1 \\ 1 & -1].
\end{align}

Next, we consider arithmetic addition and subtraction involved in the calculation of an objective function value $E(\mathbf{x_0})$.
To express these operations on a quantum circuit, the phase of the value part is advanced or delayed.
To begin with, we consider the case of addition. Let $a_1 \in \mathbb{Z}$ denote the coefficient of a term in the objective function to be added and its corresponding phase parameter is defined as $\theta_1=2\pi a_1/2^m$. The phase advance is performed by a unitary gate defined as
\begin{align}\label{eq:U}
    \mathbf{U}(\theta) = \underbrace{\mathbf{R}(2^{m - 1}\theta) \otimes \cdots \otimes \mathbf{R}(2^{0}\theta)}_{m}
\end{align}
on the value part,
where the phase gate is defined as
\begin{align}
        \mathbf{R}(\theta) = \mqty[1 & 0 \\ 0 & e^{\mathrm{j}\theta}].
\end{align}
Focusing on the $k$-th qubit in the value part, the action of $\mathbf{R}(2^{m-k}\theta_1)$ results in
\begin{align}\label{eq:single_bit_operation}
    \frac{1}{\sqrt{2}}(\Ket{0} + \ket{1})
    \xrightarrow{\mathbf{R}(2^{m-k}\theta_1)}
    \frac{1}{\sqrt{2}} (\Ket{0} + e^{\mathrm{j}2^{m-k}\theta_1}\ket{1}).
\end{align}
Therefore, applying $\mathbf{U}(\theta_1)$ to the value part results in 
\begin{align}
    \frac{1}{\sqrt{2^{m}}} (\Ket{0}+\Ket{1})_{m}
    \xrightarrow{\mathbf{U}(\theta_1)}
    \frac{1}{\sqrt{2^{m}}}\bigotimes_{k=1}^{m} (\Ket{0} + e^{\mathrm{j}2^{m-k}\theta_1}\ket{1}).
    \label{eq:AyUG}
\end{align}
Afterward, we consider the case of the subtraction. Let $a_2 \in \mathbb{Z}$ denote the coefficient of a term in the objective function to be subtracted from $a_1$, i.e., the result would be $a_1-a_2$, and its corresponding phase parameter is defined as $\theta_2=2\pi a_2/2^m$. To perform the phase delay, we just apply $\mathbf{U}(-\theta_2)$ to the quantum state of \eqref{eq:AyUG}, i.e., 
\begin{equation}
\begin{split}
    &\frac{1}{\sqrt{2^{m}}}\bigotimes_{k=1}^{m} (\Ket{0} + e^{\mathrm{j}2^{m-k}\theta_1}\ket{1}) \\
    \xrightarrow{\mathbf{U}(-\theta_2)} &\frac{1}{\sqrt{2^{m}}}\bigotimes_{k=1}^{m} (\Ket{0} + e^{\mathrm{j}2^{m-k}(\theta_1-\theta_2)}\ket{1}).
\end{split}
\end{equation}

To encode the objective function value $E(\mathbf{x}_0)$, the phase of qubits is shifted by $\theta' = 2 \pi E(\mathbf{x}_0) /2^m$ using these phase shift operations above.

Finally, by applying the inverse quantum Fourier transform (IQFT) to the value part, the value of $E(\mathbf{x}_0)$ is encoded on the quantum circuit in the form of two's complement, i.e.,
\begin{align}\label{eq:IQFT}
    \begin{split}
    &\frac{1}{\sqrt{2^{m}}}\bigotimes_{k=1}^{m} (\Ket{0} + e^{\mathrm{j}2^{m-k}\theta'}\ket{1})\\
    \xrightarrow{\mathrm{IQFT}}
    &\frac{1}{\sqrt{2^{m}}}\Ket{E(\mathbf{x}_0)~\text{mod}~2^m}_{m}.
    \end{split}
\end{align}
Since $E(\mathbf{x}_0)$ is encoded in the two's complement representation, the most significant qubit in the value part represents the sign of the value.

Using the above $\mathbf{U}(\theta)$ and IQFT, the operator $\mathbf{A}$ that calculates all possible objective function values can be constructed.
Namely, $\mathbf{U}(\theta)$ is controlled by multiple qubits in the key part associated with variables in a term, which advances the phase if all the variables are 1, and other multi-controlled $\mathbf{U}(\theta)$ gates corresponding to the other term in $E(\mathbf{x})$ act on the circuit similarly.

\begin{figure}[t]
	\centering
	\includegraphics[width=1.0\hsize]{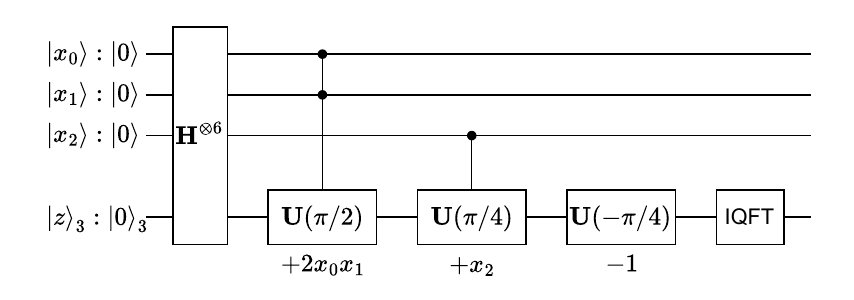}
	\caption{Quantum circuit of $\mathbf{A}$ corresponding to the objective function $E(\mathbf{x}) = 2x_0 x_1 + x_2 - 1$ with $(n, m) = (3, 3)$. Note that the notations such as $\ket{z}_3$ and $\ket{0}_3 = \ket{000}$ represent the quantum state of the 3 qubits corresponding to the value part.}
	\label{fig:conv_A}
\vspace{-.4cm}
\end{figure}
\begin{figure}[t]
	\centering
    	\includegraphics[width=1.0\hsize]{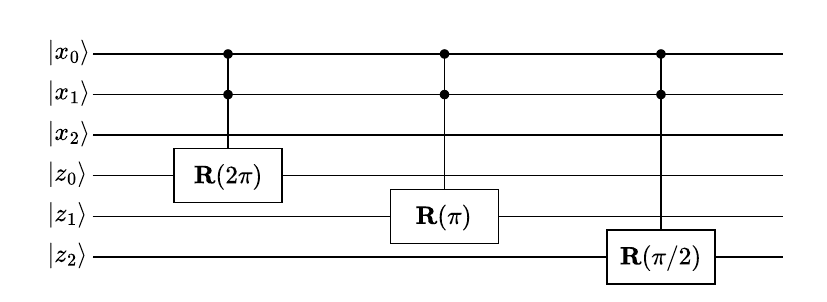}
	\caption{Components in $\mathbf{U}(\theta)$ for the term $2x_0x_1$ with $(n, m) = (3, 3)$.}
	\label{fig:conv_UG}
\vspace{-.4cm}
\end{figure}
As a concrete example, Fig.~\ref{fig:conv_A} shows the quantum circuit of $\mathbf{A}$ corresponding to the objective function $E(\mathbf{x}) = 2x_0x_1 + x_2 - 1$ with $(n,m)=(3,3)$.
As shown in Fig.~\ref{fig:conv_A}, $\mathbf{U}(\theta)$ is applied depending on each term in $E(\mathbf{x})$ and controlled by the qubits corresponding to binary variables in the term.
Additionally, Fig.~\ref{fig:conv_UG} shows the decomposition of the multi-controlled $\mathbf{U}(\theta)$ gate corresponding to the term $2x_0x_1$. 
In this case, the phase parameter is $\theta = 2\pi \cdot 2 / 2^3 = \pi/2$ and each of the three $\mathbf{R}$ gates controlled by $\ket{x_0}$ and $\ket{x_1}$ acts on a different value qubit.

To construct the multi-control $\mathbf{U}(\theta)$ gate corresponding to an $n$-th order term, it is necessary to control a phase shift gate using $n$ qubits.
The $\mathbf{R}$ gate is a $\mathcal{U}(2)$ gate.
In \cite{silva2022linear}, the decomposition method of the multi-control $\mathcal{U}(2)$ and $\mathcal{SU}(2)$ gates without using the ancilla qubit is proposed.
According to the method, to construct the multi-control $\mathcal{U}(2)$ gate with $n$ control qubits and without an ancilla qubit, $4n^2-4n+2$ CNOT gates are required.


\section{Proposed GAS with Spin Variables}
\label{sec:prop}
In this section, we propose a method to construct a GAS circuit corresponding to spin variables. By modifying the conventional construction method of Section~\ref{subsec:qd}, which corresponds to binary variables $\{0, 1\}$, we enable it to handle spin variables $\{+1, -1\}$.

In the conventional method for constructing the GAS quantum dictionary, the gate $\mathbf{U}(\theta)$ that performs the phase shift is composed of multiple multi-control phase gates $\mathbf{R}(\theta)$.
When adding a value $a \in \mathbb{R}$, the phase is advanced by applying $\mathbf{U}(\theta)$ with $\theta = a \pi / 2^{m-1}$, while in the case of subtraction, the phase is delayed by applying $\mathbf{U}(-\theta)$.
The proposed method replaces the multi-control $\mathbf{R}(\theta)$ gate with a combination of CNOT gates
\begin{align}
    \begin{bmatrix}
        1 & 0 & 0 & 0\\
        0 & 1 & 0 & 0\\
        0 & 0 & 0 & 1\\
        0 & 0 & 1 & 0
    \end{bmatrix}
\end{align}
and rotation-Z gates
\begin{align}
    \mathbf{R}_{\mathrm{z}}(\theta) = \begin{bmatrix}
        e^{-\mathrm{j}\frac{\theta}{2}} & 0 \\
        0 & e^{\mathrm{j}\frac{\theta}{2}}
    \end{bmatrix} \in \mathcal{SU}(2).
\end{align}
In the following, we describe a specific method to construct the circuit corresponding to spin variables step-by-step.

\subsection{Phase Shift Operation with $\mathbf{R}_{\mathrm{z}}$ Gate}
Here, we explain how to substitute $\mathbf{R}$ gate with $\mathbf{R}_{\mathrm{z}}$ gate maintaining the same result of the conventional gate action on the $k$-th qubit of the value part, which is given in \eqref{eq:single_bit_operation}.

\subsubsection{Gate Operation for Phase Advance}
We begin by considering the action of a rotation-Z gate that advances the phase of a single-qubit quantum state.
Acting a rotation-Z gate with the phase parameter $2^{m-k} \theta_1$ on an equal superposition state yields
\begin{align}
    \begin{split}
        \frac{1}{\sqrt{2}}(\Ket{0} + \ket{1})
        \xrightarrow{\mathbf{R}_{\mathrm{z}}(2^{m-k}\theta_1)}
        &\frac{1}{\sqrt{2}}(e^{-\mathrm{j} 2^{m-k} \frac{\theta_1}{2}}\Ket{0} + e^{\mathrm{j} 2^{m-k} \frac{\theta_1}{2}}\ket{1})\\
        =&\frac{1}{\sqrt{2}} e^{-\mathrm{j}2^{m-k} \frac{\theta_1}{2}}(\Ket{0} + e^{\mathrm{j}2^{m-k}\theta_1}\ket{1}).
    \end{split}\label{eq:phase_advance}
\end{align}
Ignoring the global phase, this result is consistent with the phase advance in the conventional quantum dictionary subroutine expressed as \eqref{eq:single_bit_operation}.
That is, in this case, the phase advance operation achieved with a phase gate $\mathbf{R}$ can be simply replaced by a rotation-Z gate $\mathbf{R}_{\mathrm{z}}$.

\subsubsection{Gate Operation for Phase Delay}
Next, we consider how to realize a gate operation with a rotation-Z gate that delays the phase of a qubit.
Let us assume that a quantum state $(\Ket{0} + e^{\mathrm{j}2^{m-k}\theta_1}\ket{1}) / \sqrt{2}$ in the value part is already created, and we aim to delay the phase by $2^{m-k}\theta_2$, so that the resulting quantum state becomes $ (\Ket{0} + e^{\mathrm{j}2^{m-k}(\theta_1-\theta_2)}\Ket{1})/\sqrt{2}$.
This operation can be realized by using a rotation-Z gate enclosed by two Pauli-X gates defined as
\begin{align}
    \mathbf{X} = \begin{bmatrix}
        0 & 1 \\
        1 & 0
    \end{bmatrix}.
\end{align}
Specifically, the transition of these actions can be described as
\begin{align}
    \begin{split}
        &\frac{1}{\sqrt{2}}(\Ket{0} + e^{\mathrm{j}2^{m-k}\theta_1}\ket{1})\\
        \xrightarrow{\mathbf{X}}
        &\frac{1}{\sqrt{2}} (e^{\mathrm{j}2^{m-k}\theta_1}\Ket{0} + \ket{1})\\
        =&\frac{1}{\sqrt{2}} e^{\mathrm{j}\alpha}(\Ket{0} + e^{-\mathrm{j}2^{m-k}\theta_1}\ket{1})\\
        \xrightarrow{\mathbf{R}_{\mathrm{z}}(2^{m-k}\theta_2)}
        &\frac{1}{\sqrt{2}} e^{\mathrm{j}\alpha}(e^{-\mathrm{j}\frac{2^{m-k}\theta_2}{2}}\Ket{0} + e^{-\mathrm{j}2^{m-k}(\theta_1 - \frac{\theta_2}{2})}\ket{1})\\
        =&\frac{1}{\sqrt{2}} e^{\mathrm{j}\beta}(\Ket{0} + e^{-\mathrm{j}2^{m-k}(\theta_1 - \theta_2)}\ket{1})\\
        \xrightarrow{\mathbf{X}}
        &\frac{1}{\sqrt{2}} e^{\mathrm{j}\beta}(e^{-\mathrm{j}2^{m-k}(\theta_1 - \theta_2)}\Ket{0} + \ket{1})\\
        =&\frac{1}{\sqrt{2}} e^{\mathrm{j}\gamma}(\Ket{0} + e^{\mathrm{j}2^{m-k}(\theta_1 - \theta_2)}\ket{1}),
    \end{split}\label{eq:phase_delay}
\end{align}
where the phases are $\alpha = 2^{m-k}\theta_1$, $\beta=2^{m-k}(\theta_1 - \frac{\theta_2}{2})$, and $\gamma= 2^{m-k-1}\theta_2$.
From this result, ignoring the global phase, the phase of the qubit is delayed by $2^{m-k}\theta_2$.

\subsection{Construction of Novel Quantum Dictionary}
Based on the above observation of the phase advance and delay operations with the $\mathbf{R}_{\mathrm{z}}$ gate, we propose a method to construct a novel quantum dictionary that calculates an objective function value using spin variables.

In the proposed method, each variable in a given term takes a value of $+1$ or $-1$, corresponding to the quantum state $\ket{0}$ or $\ket{1}$, respectively.
The phase is either advanced or delayed by an angle proportional to the term's coefficient, depending on the product of the variables.
When implementing this operation using $\mathbf{R}_{\mathrm{z}}$ gates instead of $\mathbf{R}$ gates, the operation can be realized by enclosing each $\mathbf{R}_{\mathrm{z}}$ gate between multiple CNOT gates, where the control qubits correspond to the key part’s qubits associated with the variables in the term.
These CNOT gates flip the value qubits when the state of a key qubit corresponding to a spin variable in the term is $\ket{1}$, meaning that the variable takes the value of $-1$. Conversely, no action is taken when the key qubit is in the state $\ket{0}$, meaning the variable takes the value of $+1$.
That is, the phases of the qubits in the value part are delayed only when an odd number of spin variables take a value of $-1$.

\subsection{Example of Calculation by the Proposed Quantum Dictionary}
\begin{figure*}[t]
    \centering
    \includegraphics[keepaspectratio,scale=0.7]{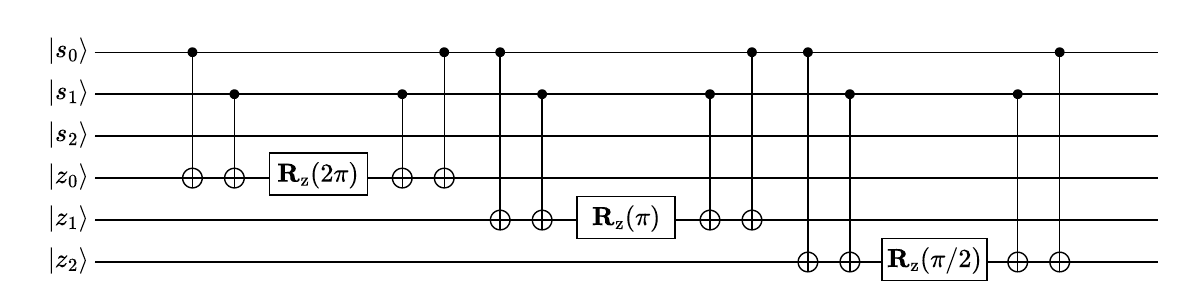}
    \caption{Proposed quantum dictionary to calculate the monomial $s_0 s_1$ with $(n,m)=(3,3)$.}
    \label{fig:prop_circuit_monomial}
\end{figure*}
As a specific example, Fig.~\ref{fig:prop_circuit_monomial} shows the quantum circuit for calculating the term $s_0 s_1$ in an objective function using spin variables, where the key part consists of $n=3$ and the value part consists of $m=3$ qubits.
Depending on the values of $s_0$ and $s_1$, their product takes the value of $+1$ or $-1$. 
An $\mathbf{R}_{\mathrm{z}}$ gate acts on the $k$-th qubit of the value part, advancing its phase by $2^{m-k}\theta$ where $\theta = 2\pi \cdot 1 /2^m = 2\pi \cdot 1 /2^3 = \pi/4$.
Each $\mathbf{R}_{\mathrm{z}}(2^{m-k}\theta)$ gate is enclosed by two CNOT gates, with the control qubits corresponding to $s_0$ and $s_1$. 
If both variables take the same value of $1$ or $-1$, the X gates that enclose the $\mathbf{R}_{\mathrm{z}}(2^{m-k}\theta)$ gate cancel out since an even number of $\mathbf{X}$ gates applied successively results in no effect, advancing the phase of the qubit by $2^{m-k}\theta$, as given in \eqref{eq:phase_advance}.

In contrast, if only one of the two variables takes a value of $-1$, the pair of X gates will enclose the $\mathbf{R}_{\mathrm{z}}(2^{m-k}\theta)$ gate, delaying the phase of the qubits by $2^{m-k}\theta$, as given in \eqref{eq:phase_delay}. This principle generalizes to a term with any number of variables. The circuit's action that advances or delays the phase depends solely on the parity of the number of spin variables equal to $-1$. An even number of variables with a value of $-1$ results in the corresponding CNOT gates canceling each other out, while an odd number results in a single X-operation enclosing the $\mathbf{R}_\mathrm{z}$ gate. 

By applying the same operation to all the qubits of the value part in the same manner as \eqref{eq:U}, the phase of qubits is shifted in proportion to the value of $s_0s_1$. Afterward, by applying IQFT as given in \eqref{eq:IQFT}, the value of the term $s_0 s_1$ can be correctly calculated.

As seen from the example, by enclosing $\mathbf{R}_{\mathrm{z}}$ gates with CNOT gates, it is possible to compute the value of an objective function with spin variables. Furthermore, to compute the $n$-th term, only $2mn$ CNOT gates and $m$ $\mathbf{R}_{\mathrm{z}}$ gates are required, eliminating the need for multi-controlled phase gates used in the circuit design of the conventional quantum dictionary subroutine.

\begin{figure}[t]
    \centering
    \begingroup
    \subfigure[Binary-based calculation]{
        \centering
        \includegraphics[width=0.45\linewidth]{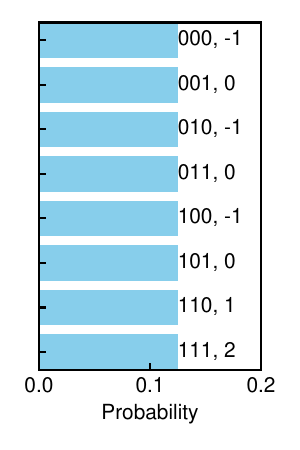}
        \label{fig:conventional_Ay}
    }
    \hfill
    \subfigure[Spin-based calculation]{
        \centering
        \includegraphics[width=0.45\linewidth]{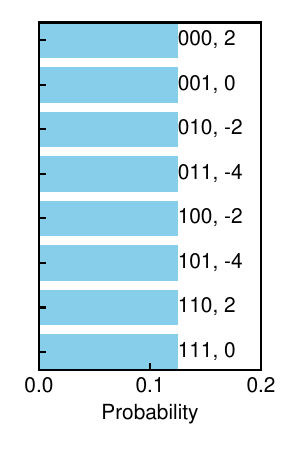}
        \label{fig:proposed_Ay}
    }
    \caption{Comparison of the probabilities that quantum states are observed after applying the conventional (left) and proposed (right) quantum subroutines.}\label{fig:Ay_comparison}
    \endgroup
    \vspace{-.4cm}
\end{figure}

In addition, in Fig.~\ref{fig:Ay_comparison}, we show the results of the calculations of $E(\mathbf{x})=2x_0x_1 + x_2 - 1$ and $E(\mathbf{s})=2s_0s_1 + s_2 - 1$ using the conventional and proposed quantum dictionary designs.
In Figs.~\ref{fig:Ay_comparison}(a) and (b), three binary digits next to each bar represent the state of qubits in the key part.
Note that in the proposed method, when the state of each qubit is $\ket{0}$ or $\ket{1}$, it expresses $s = +1$ or $s = -1$ in the objective function.
From this result, it can be seen that different values are encoded, and the proposed quantum dictionary designs can calculate the value of the objective function with spin variables correctly.


\section{Gate Reduction on Practical Use Cases}
\label{sec:gate_reduction}
In this section, we consider two practical use cases of the GAS algorithm, and algebraically analyze the numbers of terms in their objective functions and CNOT gates required to construct a quantum dictionary subroutine, demonstrating the advantage of the proposed circuit design in terms of gate reduction.

\subsection{Use Case 1: MIMO ML Detection}\label{subsec:MIMO_detection}
In modern wireless communication standards, multiple-input multiple-output~(MIMO) systems have become an essential key technology and been widely adopted to achieve enhanced communication throughput and quality, which have an inevitable trade-off.
In MIMO systems, symbols conveying information are transmitted from $N_{\mathrm{t}}$ antennas and received by $N_{\mathrm{r}}$ antennas, where $N_{\mathrm{t}}, N_{\mathrm{r}} \geq 2$.
The received symbol vector $\mathbf{r} \in \mathbb{C}^{N_{\mathrm{r}}\times 1}$ is expressed as
\begin{align}
	\mathbf{r} = \frac{1}{\sqrt{N_{\mathrm{t}}}}\mathbf{H}_{\mathrm{c}}\mathbf{t} + \sigma \mathbf{n},
\end{align}
where $\mathbf{H}_{\mathrm{c}} \in \mathbb{C}^{N_{\mathrm{r}}\times N_{\mathrm{t}}}$ is a channel matrix,
$\mathbf{t} \in \mathbb{C}^{N_{\mathrm{t}}\times 1}$ is a vector containing $N_{\mathrm{t}}$ symbols drawn from a constellation $\mathcal{C}$,
and $\mathbf{n} \in \mathbb{C}^{N_{\mathrm{r}}\times 1}$ is a vector of additive white Gaussian noise.
Each element of $\mathbf{H}_{\mathrm{c}} $ and $\mathbf{n}$ follow the standard complex Gaussian distribution $\mathcal{CN}(0,1)$.
The goal of MIMO detection is to estimate the transmitted symbol vector $\mathbf{t}$ based on the received symbol vector.
The ML detection is the optimal method for estimating the transmitted symbol vector by minimizing the Euclidean distance between the received and candidate symbol vectors, i.e., 
\begin{align}
	\hat{\mathbf{t}} = \argmin_{\mathbf{t}\in \mathcal{C}^{N_{\mathrm{t}}}} \left\|\mathbf{r} - \frac{1}{\sqrt{N_{\mathrm{t}}}}\mathbf{H}_{\mathrm{c}}\mathbf{t}\right\|^2.
\end{align}

The ML detection is a combinatorial optimization problem that requires an exhaustive search over all possible transmitted symbol vectors.
This problem is one instance of the multi-user detection problem, which is classified as NP-hard \cite{verdu1989computationa}. 
An approximate solution to this problem can be obtained through various approaches, such as the semidefinite relaxation technique \cite{luo2010semidefinite}, a genetic algorithm, and a simulated annealing algorithm \cite{Abrasao2010MIMO}. 
These approaches can find the approximate solution with lower computational complexity compared to that required by the ML detection. 
In practical systems, the detection is often carried out with a symbol-by-symbol basis where a linear equalizer cancels interference to provide a feasible solution without the need for exhaustive search.
However, they do not achieve the optimal performance provided by the ML detection, indicating a trade-off between performance and complexity for detection.

The approach to solve the ML detection problem with the GAS algorithm has been proposed in \cite{norimoto2023quantum}.
This application achieves a quadratic speedup compared to a classical exhaustive search,
which indicates the potential of the GAS algorithm for future communication systems.
In this conventional approach, the ML detection problem is formulated as a HUBO problem,
whose objective function is the distance between the received and candidate symbol vectors, i.e.,
\begin{equation}
    \begin{aligned}
 \min_{\mathbf{x}} &\quad E(\mathbf{x})= \left\|\mathbf{r}-\frac{1}{\sqrt{N_{\mathrm{t}}}}\mathbf{H}_{\mathrm{c}} \mathbf{t} (\mathbf{x})\right\|^{2}\\
 \textrm{s.t.}& \quad x_i \in \{0, 1\} ~~~~ (i = 0, \cdots, 2MN_{\mathrm{t}} -1),
\label{eq:MIMO_objfun_binary}
    \end{aligned}
\end{equation}
where $\mathbf{t} (\mathbf{x}) = [t_0(\mathbf{x}),\cdots, t_{N_{\mathrm{t}}-1} (\mathbf{x}) ]^{\mathrm{T}}$ and each element of the vector is a function of $\mathbf{x}$.
In this formulation, each transmitted symbol is modulated by the well-known Gray labeling method,
which is adopted in the 5G standard~\cite{3gpp2018ts}.
Based on the equations for binary phase shift keying~(BPSK), quadrature phase shift keying~(QPSK) and $16$-quadrature amplitude modulation~(QAM) specified in \cite{3gpp2018ts}, a general equation expressing a $2^{2M}$-QAM symbol, where $M$ is a positive integer, can be described as 
\begin{align}
\label{eq:binary-map}
t_v(\mathbf{x})=&\frac{1}{\sqrt{A}}\sum_{k=0}^{M-1}(-1)^{k}\cdot 2^{M-1-k} \prod_{l=0}^{k}(1-2x_{2Mv+2l}) \notag \\
&+\frac{\mathrm{j}}{\sqrt{A}}\sum_{k=0}^{M-1}(-1)^{k}\cdot 2^{M-1-k} \prod_{l=0}^{k}(1-2x_{2Mv+(2l+1)}),
\end{align}
where $t_v(\mathbf{x})$ denotes the $v$-th element of the vector $\mathbf{t}(\mathbf{x})$ and the normalization factor $A$ is calculated as 
\begin{align}
    A = \frac{1}{2^{M-1}} \sum_{k=0}^{2^M-1} \left(2k - 2^M + 1\right)^2.
\end{align}

This modulation method results in a product of $2M$ brackets in the form of $(1-2x)$ and 
its expansion leads to $2^{2M}$ terms in the objective function, which causes exponential growth of the circuit for $\mathbf{A}_y$.

To avoid this issue, we consider the reformulation of the problem utilizing spin variables, i.e., 
\begin{equation}
    \begin{aligned}
 \min_{\mathbf{s}} & \quad E(\mathbf{s})=  \left\|\mathbf{r}-\frac{1}{\sqrt{N_{\mathrm{t}}}}\mathbf{H}_{\mathrm{c}} \mathbf{t} '(\mathbf{s})\right\|^{2}\\
 \textrm{s.t.} &\quad s_i \in \{+1, -1\} ~~~~ (i = 0, \cdots, 2MN_{\mathrm{t}} -1),
\label{eq:MIMO_objfun_spin}
    \end{aligned}
\end{equation}
where $\mathbf{t}' (\mathbf{s}) = [t'_0(\mathbf{s}),\cdots, t'_{N_{\mathrm{t}}-1} (\mathbf{s}) ]^{\mathrm{T}}$.
Each symbol $t'_v(\mathbf{s})$ for $v = 0, \cdots, N_{\mathrm{t}}-1$ is obtained by replacing each bracket in the form of $(1-2x)$ in \eqref{eq:binary-map} with a spin variable, and can be written as
\begin{equation}
\begin{aligned}
    \label{eq:map-qam}
t'_v(\mathbf{s})=&\frac{1}{\sqrt{A}}\sum_{k=0}^{M-1}(-1)^{k}\cdot 2^{M-1-k} \prod_{l=0}^{k} s_{2Mv+2l}\\ 
&+\frac{\mathrm{j}}{\sqrt{A}}\sum_{k=0}^{M-1}(-1)^{k}\cdot 2^{M-1-k} \prod_{l=0}^{k}s_{2Mv+(2l+1)},
\end{aligned}
\end{equation}
which simplifies the objective function, as each product of spin variables results in a single monomial, preventing exponential growth in the number of terms.

In the following, we compare the numbers of terms in the objective functions with binary and spin variables.
Let $N^{\mathbf{b}}_{k}$ denote the number of $k$-th order terms in the objective function with binary variables.
This number is expressed for $1 \leq k \leq 2M$ as
\begin{equation}\label{eq:nrof_terms_binary}
N_{k}^{\mathbf{b}}=\begin{cases}\binom{2M}{k}2N_{\mathrm{t}}(N_{t}-1)-\binom{M}{k}2N_{t}(2N_{t}-3) & \text{if } k \le M \\ \binom{2M}{k}2N_{\mathrm{t}}(N_{t}-1)& \text{if } k > M. \end{cases}
\end{equation}
Meanwhile, 
let $N^{\mathbf{s}}_{k}$ denote the number of $k$-th order terms in the objective function with spin variables.
This number is expressed for $1 \leq k \leq 2M$ as
\begin{align}\label{eq:nrof_terms_spin}
N_k^{\mathbf{s}} =
\begin{cases}
    (k-1) 2N_{\mathrm{t}}(N_{\mathrm{t}}-1) + (M-k + 1)2N_{\mathrm{t}} & \text{if } k \le M, \\
    (2M-k+1) 2N_{\mathrm{t}}(N_{\mathrm{t}}-1) & \text{if } k > M.
\end{cases}
\end{align}

The detailed derivation of these formulas is described in Appendix.
The total number of terms in the objective function with binary variables is calculated as the summation of $N^{\mathbf{b}}_{k}$ over $1 \leq k \leq 2M$, i.e., 
\begin{equation}
\begin{split}
    \sum_{k=1}^{2M} N^{\mathbf{b}}_{k} &= (2^{2M}-1)\cdot2N_{\mathrm{t}}(N_{\mathrm{t}}-1) \\
    &\ \ \ \ -  (2^{M} -1)\cdot 2N_{\mathrm{t}}(2N_{\mathrm{t}}-3) \\
    &= O(2^{2M} N_{\mathrm{t}}^2).
\end{split}
\end{equation}
Similarly, for the spin-based case, the total number is calculated as
\begin{equation}
\begin{split}
    \sum_{k=1}^{2M} N^{\mathbf{s}}_{k} &=  M^2 \cdot 2N_{\mathrm{t}}(N_{\mathrm{t}}-1)\\ 
    &\ +   \frac{M(M+1)}{2}\cdot 2N_\mathrm{t}\\
    &= O(M^2N_{\mathrm{t}}^2).
\end{split}
\end{equation}
These results indicate that the total number of terms can be significantly reduced from an exponential order  $O(2^{2M} N_{\mathrm{t}}^2)$ to a polynomial order $O(M^2 N_{\mathrm{t}}^2)$ when the function is formulated with spin variables, which leads to an improvement of the feasibility to construct the circuit for $\mathbf{A}_y$. 

\subsection{Use Case 2: Syndrome Decoding Problem}\label{subsec:syndrome_decoding}
A syndrome decoding problem has been discussed in coding theory and information theory.
Its NP-hardness has been proven in \cite{berlekamp1978inherent} and guarantees the security of cryptography such as McEliece cryptosystem \cite{mceliece1978public}. 

Let $\mathbf{H}_{\mathrm{p}} \in \mathbb{F}^{M\times N}_2$ denote a parity check matrix,
where $N$ is the length of codeword $\mathbf{x} \in \mathbb{F}^{N}_2$ and $M$ is the number of parity check bits.
The goal of the syndrome decoding problem is to find 
$\mathbf{x}$  with a Hamming weight less than or equal to $w~(\leq N)$ that satisfies 
\begin{align}
     \mathbf{H}_{\mathrm{p}}\mathbf{x} = \mathbf{y},\label{eq:syndrome_eq}
\end{align}
where $\mathbf{y} \in \mathbb{F}^{M}_2$ is a syndrome. 
If $\mathbf{x}$ does not satisfy that equation, it is considered as an error occurred.

To solve this problem with the GAS algorithm, it is required to design an objective function that gives a minimum value when the input vector $\mathbf{x} $ satisfies \eqref{eq:syndrome_eq}. 
Specifically, the objective function value increases when each element of $\mathbf{H}_{\mathrm{p}}\mathbf{x}$ and $\mathbf{y}$ does not coincide.
That is, the problem suitable for the GAS algorithm with the conventional circuit design is formulated as 
\begin{equation}
\begin{aligned}\label{eq:conv_syndrome}
    \min_{\mathbf{x}} &\quad E(\mathbf{x}) = -\left(\sum_{j=0}^{M-1} (-1)^{y_j} \prod_{\substack{i=0}}^{N-1}(1-2x_i)^{h_{j,i}}\right) \\
     \textrm{s.t.} &\quad x_i \in \{0, 1\} ~~~~ (i = 0, \cdots, N-1 ),
\end{aligned}
\end{equation}
where $h_{j,i}$ is the element at $j$-th row and $i$-th column of the matrix $\mathbf{H}_{\mathrm{p}}$. 
In the objective function, XOR operation between values on $\mathbb{F}_2$ is expressed as a product of brackets in the form of $(1-2x)$.
If that product takes a value of $+1$, it indicates that the XOR operation results in $0$ on $\mathbb{F}_2$, whereas if it takes a value of $-1$, it indicates the XOR operation results in $1$.
Note that constraint terms should be added to the equation above to ensure that the solution $\mathbf{x}$ has a Hamming weight not exceeding $w$.
However, that constraint can be omitted when the initial state is prepared as the superposition of quantum states whose Hamming weights are less than or equal to $w$, which can be prepared as a symmetric pure state \cite{bartschi2019deterministic}.
This formulation exponentially increases the number of terms in the objective function to the order of $O(2^N)$ in the worst case.

In contrast, the objective function can be simplified by reformulation with spin variables.
Each bracket $(1-2x)$ in \eqref{eq:conv_syndrome} is substituted with a spin variable, which is similar to the one in \eqref{eq:map-qam}.
Then, the reformulation of the problem that is suitable for the GAS algorithm with the proposed circuit design is expressed as 
\begin{equation}
\begin{aligned}\label{eq:prop_syndrome}
    \min_{\mathbf{s}} &\quad E(\mathbf{s}) = -\left(\sum_{j=0}^{M-1}(-1)^{y_j}\prod_{\substack{i=0}}^{N-1}s_i^{h_{j,i}}\right)\\
     \textrm{s.t.} &\quad s_i \in \{+1, -1\} ~~~~ (i = 0, \cdots, N-1 ).
\end{aligned}
\end{equation}
This formulation results in the reduced number of terms of order $O(M)$ in the objective function even in the worst case.

\section{Performance Analysis}
\label{sec:performance}
In this section, we compare the conventional and proposed circuit designs in terms of query complexity of the GAS algorithm and the number of CNOT gates required to construct a quantum circuit for $\mathbf{A}_y$, considering the problems of the two practical use cases introduced in Section~\ref{sec:gate_reduction}. 
In the complexity comparisons, two types of metrics are considered: complexities in the classical domain (CD) and the quantum domain (QD), which are used in \cite{botsinis2014fixedcomplexity}. The complexity in CD is defined as the number of measurements of quantum states, i.e., it corresponds to $c$, while the complexity in QD is defined as the total number of Grover operators $\mathbf{G}$, i.e., $L_0 + L_1 + \cdots + L_c$, which is termed query complexity. $L_c$ is updated at each iteration as described in the GAS algorithm explanation in Section II-A.
The complexity in CD has not been widely considered in the context of quantum computing, but holds significance from a practical perspective. 
Furthermore, the GAS simulation was performed based on the mathematical equations describing the ideal behavior, under the assumption that the number of qubits in the value part, $m$, is sufficiently large.
In addition, we also simulated the performance of classical exhaustive search for the reference, by counting the number of objective function evaluations. This metric serves as a classical analogue to the CD complexity, providing a baseline for performance comparison.

\subsection{MIMO ML Detection}

We compared the conventional and proposed circuit designs for the ML detection of an $N_{\mathrm{t}}=N_{\mathrm{r}}=2$ MIMO system in $16$-QAM case.
In the following simulations,
the channel matrix $\mathbf{H}_{\mathrm{c}}$ was fixed to 
\begin{equation}
\mathbf{H}_{\mathrm{c}} =
    \begin{bmatrix}
       0.749 - 0.0149\mathrm{j}& 1.32 + 0.0630\mathrm{j}\\
        0.637 - 0.143\mathrm{j} & -0.389 - 0.152\mathrm{j}
    \end{bmatrix}. \label{eq:exHc}
\end{equation}
The transmit symbol vector $\mathbf{t}$ was fixed to
\begin{align}
    \mathbf{t} = \begin{bmatrix}
        \frac{1}{\sqrt{10}}(1 + \mathrm{j}), &  \frac{1}{\sqrt{10}}(-3 - 3\mathrm{j})
    \end{bmatrix}^{\mathrm{T}}
\end{align}
corresponding to $\mathbf{x} = [0,0,0,0,1,1,1,1]^{\mathrm{T}}$.
The signal-to-noise ratio (SNR) was fixed to $20$dB corresponding to $\sigma = 0.1$ in both cases.

\begin{figure}[!t]
	\centering
	\includegraphics[width=1.0\linewidth]{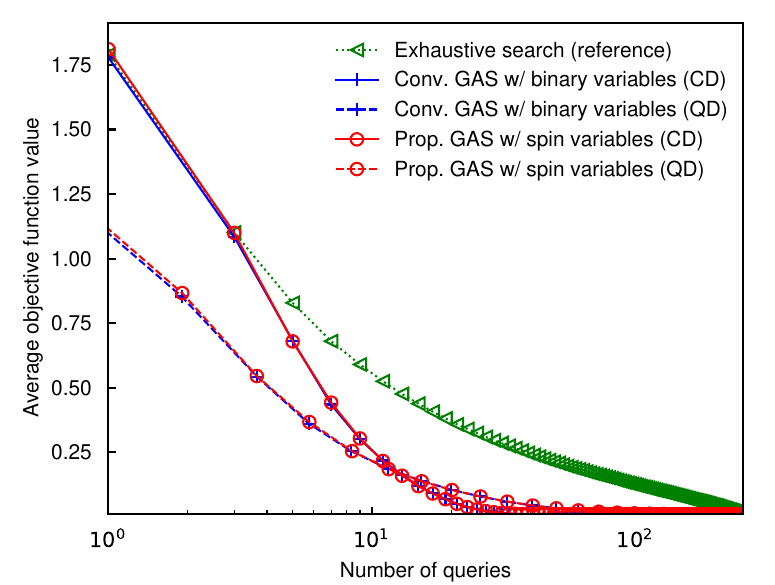}
        \label{fig:MIMO_16QAM_objfun}
    \caption{Relationship between query complexity and the average objective function values for MIMO ML detection, where $16$-QAM was adopted.}
    \label{fig:MIMO_objfun}
	\vspace{-.4cm}
\end{figure}

\begin{figure}[!t]
	\centering
        \includegraphics[width=1.0\linewidth]{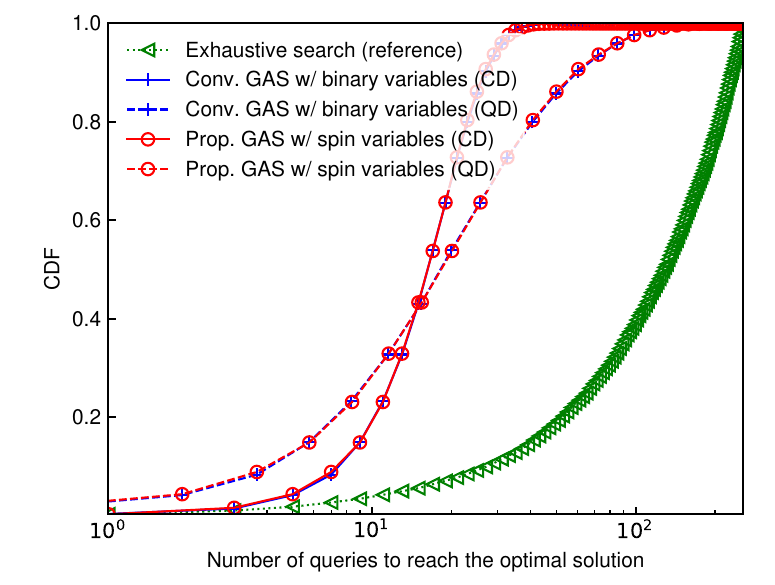}
        \label{fig:MIMO_16QAM_CDF}
    \caption{CDF of the number of queries required to reach an optimal solution of MIMO ML detection, where $16$-QAM was adopted.}
    \label{fig:MIMO_CDF}
	\vspace{-.4cm}
\end{figure}

Firstly, Fig.~\ref{fig:MIMO_objfun} shows the relationship between query complexity and average of objective function values.  
As shown in the figure, the circuit design had no effects on the convergence performance or rate.
Additionally, in Fig.~\ref{fig:MIMO_CDF}, the cumulative distribution functions (CDFs) of the number of queries to reach the optimal solution were compared.
The figure 
also indicates that the GAS algorithm with proposed circuit design achieved the same convergence performance of the objective function values as that of the conventional design.
From the results of Fig.~\ref{fig:MIMO_objfun} and Fig.~\ref{fig:MIMO_CDF}, we can observe that the GAS algorithm with both conventional and proposed circuit designs achieved the same quantum speedup of the query complexity compared to the classical exhaustive search in the case of $16$-QAM.

\begin{figure}[!t]
	\centering
	\includegraphics[width=1.0\hsize]{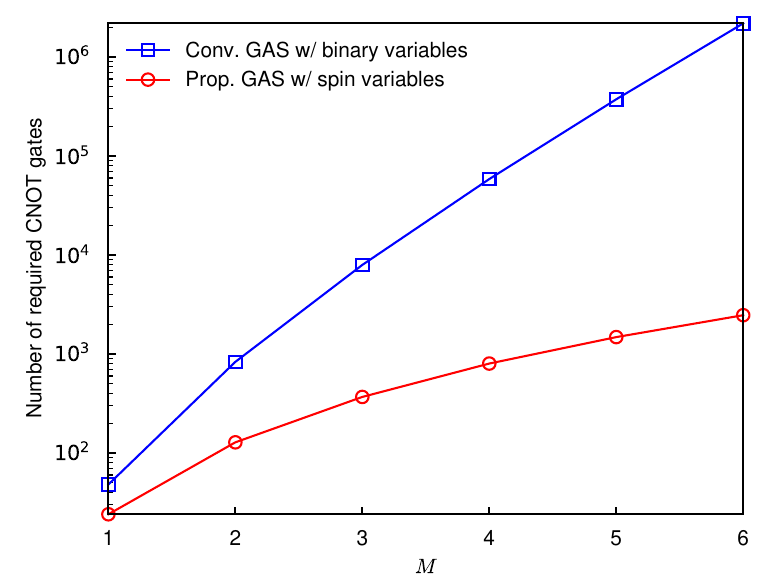}
	\caption{Number of CNOT gates required to decompose the quantum circuit of conventional and proposed $\mathbf{A}_y$ for MIMO ML detection where each symbol is modulated as $2^{2M}$-QAM symbol. Note that Y-axis is on a logarithmic scale.}
	\label{fig:MIMO_gate_count}
	\vspace{-.4cm}
\end{figure}

Secondly, in Fig.~\ref{fig:MIMO_gate_count}, we compared the number of CNOT gates per value qubit required to construct the quantum circuit for $\mathbf{A}_y$, without IQFT, in both the conventional and proposed designs of the quantum dictionary subroutine. 
The graph shows results for $M$ values ranging from 1 to 6, corresponding to QAM constellations from QPSK ($2^{2\cdot 1}$-QAM) to 4096-QAM ($2^{2\cdot 6}$-QAM).
The numbers of $n$-th order terms in the objective functions with binary and spin variables
can be theoretically evaluated
by \eqref{eq:nrof_terms_binary} and \eqref{eq:nrof_terms_spin},
respectively.
Furthermore, an $n$-controlled phase gate can be decomposed with
$4 n^2 - 4n + 2$ CNOT gates by the state-of-the-art method of \cite{Silva2023LinearDecomposition}.
Then, the number of CNOT gates required for the conventional circuit design was evaluated as the summation of the product between the number of $n$-th terms and $4 n^2 - 4n + 2$ over all possible $n$. 
Meanwhile, the number of CNOT gates required for the proposed design was evaluated as the summation of the product between the number of $n$-th terms and $2n$ over all possible $n$. 
Fig.~\ref{fig:MIMO_gate_count} clearly demonstrates that the number of the CNOT gates required for the proposed circuit design scaled in a polynomial order with respect to 
$M$, in contrast to the conventional circuit design, which exhibited an exponential growth in the number of the CNOT gates. Note that the y-axis of the figure is plotted on a logarithmic scale.
This result highlights a significant improvement, as the proposed approach reduces the depth of the circuit for $\mathbf{A}_y$.

\subsection{Syndrome Decoding Problem}
In this subsection, we compare the conventional and proposed circuit designs for the syndrome decoding problem. 
In the following, we assume $w=N$ and $\mathbf{y} =\mathbf{0}$ for the simplicity of the simulations.

As a parity check matrix of the Hamming code, we used a matrix
\begin{align}
    \mathbf{H}_{\mathrm{p}}^{(7,4)}=\begin{bmatrix}1&0&0&1&1&1&0\\
0&1&0&0&1&1&1\\
0&0&1&1&1&0&1\\
\end{bmatrix}.\label{eq:hamming-parity}
\end{align}
By increasing the number of rows and columns of the matrix by one each,
and filling the bottom row with ones,
a parity matrix of its extended Hamming code is obtained as 

\begin{align}
    \mathbf{H}_{\mathrm{p}}^{(8,4)}=\begin{bmatrix}1&0&0&1&1&1&0&0\\
0&1&0&0&1&1&1&0\\
0&0&1&1&1&0&1&0\\
1&1&1&1&1&1&1&1\\
\end{bmatrix}.\label{eq:extended-hamming-parity}
\end{align}

\begin{figure}[!t]
	\centering
        \includegraphics[width=1.0\linewidth]{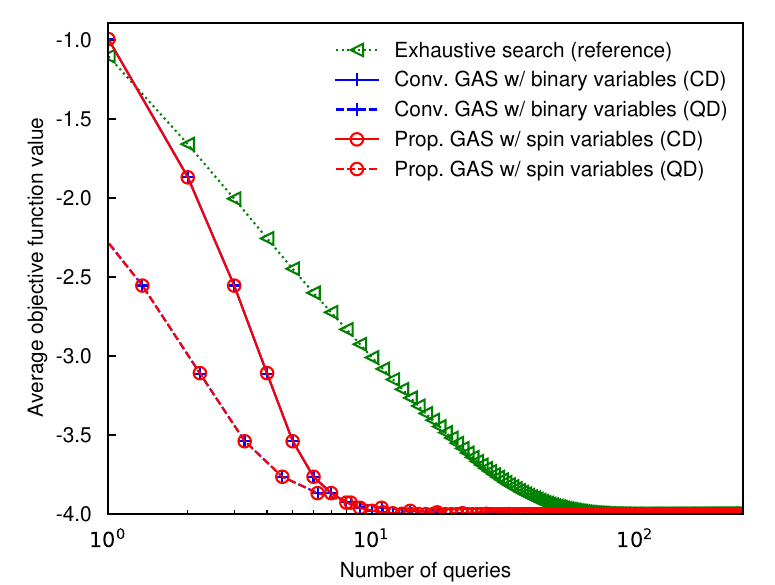}
        \label{fig:eHamming_meanobj}
    \caption{Relationship between query complexity and the average objective function values for syndrome decoding problem, where target parity check matrix was  $\mathbf{H}_{\mathrm{p}}^{(8,4)}$.}
    \label{fig:syndrome_decoding_objfun}
	\vspace{-.4cm}
\end{figure}

\begin{figure}[!t]
	\centering
        \includegraphics[width=1.0\linewidth]{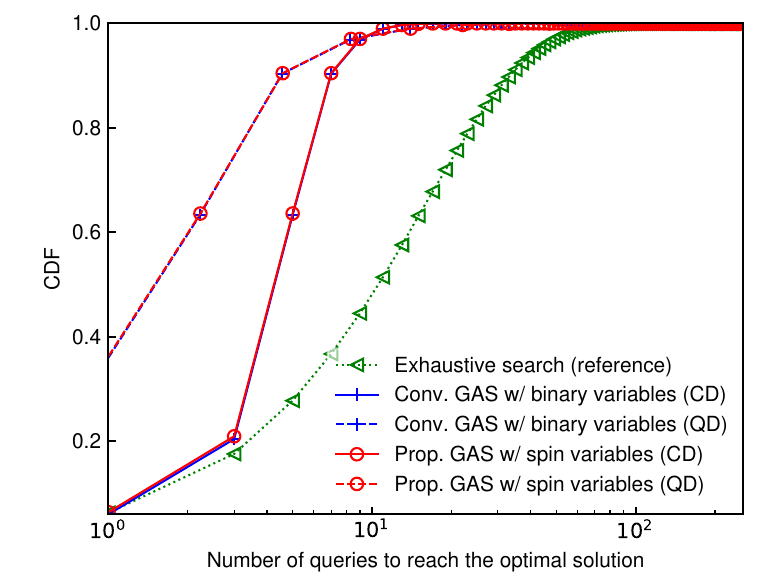}
        \label{fig:eHamming_CDF}
    \caption{CDF of the number of queries required to reach an optimal solution of syndrome decoding problem, where target parity check matrix was $\mathbf{H}_{\mathrm{p}}^{(8,4)}$.}
    \label{fig:syndrome_decoding_CDF}
	\vspace{-.4cm}
\end{figure}

Thirdly, Fig.~\ref{fig:syndrome_decoding_objfun} shows the relationship between query complexity and average of the objective function values in the case that the extended Hamming code \eqref{eq:extended-hamming-parity} is applied.
From the figure, it can be seen that the objective function values for the proposed circuit design converged in the same way as that for the conventional circuit design
in both CD and QD, exhibiting a speedup of query complexity compared to that of the classical exhaustive search.
Furthermore, Fig.~\ref{fig:syndrome_decoding_CDF} shows the CDFs of the number of queries to reach an optimal solution in the case that the extended Hamming code \eqref{eq:extended-hamming-parity} is applied.
The result indicates that the GAS algorithm with both conventional and proposed circuit designs resulted 
in the same convergence performance in the case of the extended Hamming code.

Finally, Table~\ref{table:syndrome_decoding_terms} compares the number of terms in the objective functions of \eqref{eq:conv_syndrome} and \eqref{eq:prop_syndrome}. 
It shows that, with the proposed circuit design, the number of terms for the Hamming code was reduced by $1 - 3/38 \approx 92.1\%$, while for the extended Hamming code, it was reduced by $1 - 4/256 \approx 98.4\%$. 
Similarly, Table~\ref{table:syndrome_decoding_CNOT} presents the comparison of the CNOT gates per one value qubit required to construct the circuit of $\mathbf{A}_y$ without IQFT, demonstrating a reduction by $1 - 24/626 \approx 96.2\%$ for the Hamming code and $1 - 40/14846 \approx 99.7\%$ for the extended Hamming code. 
These results indicate that the proposed circuit design offered a significant advantage, particularly when the rows of the parity check matrix contained a relatively larger number of ones.

\begin{table}[t]
\centering
\caption{Comparison of the number of terms in the syndrome decoding problem.\label{table:syndrome_decoding_terms}}
\begin{tabular}{ccc}
\multicolumn{3}{c}{}\\ \hline
\textbf{Parity Check Matrix} & \textbf{Conventional} & \textbf{Proposed} \\
\hline
$\mathbf{H}_{\mathrm{p}}^{(7,4)}$ & 38 & 3 \\
\hline
$\mathbf{H}_{\mathrm{p}}^{(8,4)}$& 256 & 4 \\
\hline
\end{tabular}
\end{table}

\begin{table}[t]
\centering
\caption{Comparison of the number of CNOT gates required in the syndrome decoding problem.\label{table:syndrome_decoding_CNOT}}
\begin{tabular}{ccc}
\multicolumn{3}{c}{}\\ \hline
\textbf{Parity Check Matrix} & \textbf{Conventional} & \textbf{Proposed} \\
\hline
$\mathbf{H}_{\mathrm{p}}^{(7,4)}$ & 626  & 24 \\
\hline
$\mathbf{H}_{\mathrm{p}}^{(8,4)}$& 14846 & 40 \\\hline
\end{tabular}
\end{table}


\section{Conclusions}
\label{sec:conc}
In this paper, we proposed a new approach to GAS that handles objective functions using spin variables instead of binary variables. 
Specifically, we introduced a method for constructing a quantum dictionary capable of efficiently calculating the values of the objective function. 
Furthermore, we demonstrated that by reformulating objective functions with spin variables, the number of terms and the required CNOT gates for constructing the quantum dictionary circuit can be reduced from an exponential order to a polynomial order, especially in applications such as the ML detection problem for MIMO systems and solving the syndrome decoding problem with the GAS algorithm.
These applications highlight the effectiveness of our proposed circuit design for problems involving XOR operations, commonly found in coding theory. As future work, we plan to explore broader applications of this approach to other problem domains, as well as investigate further reductions in the number of required CNOT gates.

\section*{Acknowledgement}
The authors would like to thank Ryuhei Mori, Nagoya University, Japan, for providing a comment on the replacement of the phase gate.

\appendix
\label{sec:appendix}
\subsection{Number of Terms in the Objective Functions for MIMO ML Detection with GAS}
We derive the general expression for the number of terms in the objective function for MIMO ML detection with GAS.
Each symbol in $\mathbf{t}$ is denoted as $t_v(\mathbf{x}) = t_v$ for the simplicity of notation.

The objective function of \eqref{eq:MIMO_objfun_binary} can be rewritten as 
\begin{align}
    E(\mathbf{x}) = \frac{1}{\sqrt{N_{\mathrm{t}}}} \sum_{u=0}^{N_{\mathrm{r}}-1} \left| r_u - \sum_{v=0}^{N_{\mathrm{t}} - 1} h_{uv} t_{v} \right| \left| r_u^{*} - \sum_{v=0}^{N_{\mathrm{t}} - 1}  h_{uv}^{*} t_{v}^{*} \right|.\label{eq:MIMO_objfun}
\end{align}
Since we focus on the number of terms that appear in the expansion of the above equation, we ignore the coefficient $1/\sqrt{N_{\mathrm{t}}}$. 
Additionally, the number of receiver antennas $N_{\mathrm{r}}$ does not affect the number of the terms in \eqref{eq:MIMO_objfun}. 
Then, we assume $N_{\mathrm{r}} = 1$ in the following analysis.
Under these assumptions, the objective function is expanded as \eqref{eq:mimo_objfun_expansion}, where $t_{v,R}$ and $t_{v,I}$ denote the real and imaginary parts of the transmitted symbol $t_v$, respectively. 
\begin{figure*}[t!]
	\begin{align}
	E(\mathbf{x}) = & \left| r_u - \sum_{v=0}^{N_{\mathrm{t}} - 1} h_{uv} t_{v} \right| \left| r_u^{*} - \sum_{v=0}^{N_{\mathrm{t}} - 1} h_{uv}^{*} t_{v}^{*} \right|\notag\\
=& |r_u|^{2} - \sum_{v=0}^{N_{\mathrm{t}} - 1} r_u^{*} h_{uv} t_{v} - \sum_{v=0}^{N_{\mathrm{t}} - 1} r_u h_{uv}^{*} t_{v}^{*} + \sum_{v=0}^{N_{\mathrm{t}} - 1} \sum_{v'=0}^{N_{\mathrm{t}} - 1} h_{uv} h_{uv'}^{*} t_v t_{v'}^{*} \notag\\	
 =& \underbrace{|r_u|^{2}}_{\text{First block}} - \underbrace{\sum_{v=0}^{N_{\mathrm{t}} - 1} 2 \text{Re}[r_u^{*} h_{uv} t_v]}_{\text{Second block}} + \underbrace{\sum_{v=0}^{N_{\mathrm{t}} - 1} |h_{uv}|^2 \left(t_{v,R}^2 + t_{v,I}^2\right)}_{\text{Third block}} \notag \\
 &+ \underbrace{2 \sum_{v=0}^{N_{\mathrm{t}}-2} \sum_{v'=v+1}^{N_{\mathrm{t}}-1} \left\{ \text{Re}(h_{uv} h_{uv'}^{*}) \left(t_{v,R} t_{v',R} + t_{v,I} t_{v',I}\right) + \text{Im}(h_{uv} h_{uv'}^{*}) \left(t_{v,R} t_{v',I} - t_{v,I} t_{v',R}\right) \right\}}_{\text{Fourth block}} \label{eq:mimo_objfun_expansion}
	\end{align}
	\hrulefill
	\vspace*{4pt}
\end{figure*}

First, we consider the case with binary variables, $N_k^b$. The total number of unique $k$-th order terms is calculated using a constructive approach by partitioning all possible terms into two disjoint sets: terms depending on variables from two distinct transmitted symbols (set $\mathrm{A}$) and terms depending on variables from only a single transmitted symbol (set $\mathrm{B}$). The total count is therefore $N_k^b = |\mathrm{A}| + |\mathrm{B}|$. The detailed derivation of $|\mathrm{A}|$ and $|\mathrm{B}|$ is explained below.

For high-order terms ($M+1 \leq k \leq 2M$), 
we begin with calculating $|\mathrm{A}|$ by analyzing the expansion of the fourth block of \eqref{eq:mimo_objfun_expansion}. Now, let us select a pair of indices $(v, v')$ from all possible $N_\mathrm{t}(N_\mathrm{t}-1)/2$ pairs of indices. 
This pair generates four products of components, $(t_{v,R} t_{v',R})$, $(t_{v,R} t_{v',I})$, $(t_{v,I} t_{v',R})$, and $(t_{v,I} t_{v',I})$. 
To calculate the number of terms generated from these products, we first consider one product, such as $t_{v,R}t_{v',R}$. The term $t_{v,R}$ is a polynomial in $M$ distinct variables, and $t_{v',R}$ is a polynomial over another disjoint set of $M$ variables. 
Their product results in the products of $2M$ brackets that are in the form of $(1-2x)$, where all $2M$ variables are distinct. Therefore, a $k$-th order term is formed by selecting a product of $k$ of these variables, and the number of ways to do so is given by the binomial coefficient $\binom{2M}{k}$.
Since there are four product types for each of the $\frac{N_t(N_t-1)}{2}$ pairs, the total count of all generated terms is $4 \cdot \binom{2M}{k} \cdot \frac{N_t(N_t-1)}{2} = \binom{2M}{k}2N_t(N_t-1)$.
Note that these terms must contain variables from at least two different transmitted symbols, meaning all unique terms belong to set $\mathrm{A}$, and set $\mathrm{B}$ is empty ($|\mathrm{B}|=0$) since one transmitted symbol only contains $M$ variables. 
Therefore, the total count is $|\mathrm{A}| + |\mathrm{B}| = \binom{2M}{k}2N_t(N_t-1)$
 
For low-order terms ($1 \leq k \leq M$), we begin with counting the number of terms in $|\mathrm{A}|$ as well. In the similar manner as the high-order case, we firstly assume that $|\mathrm{A}| = \binom{2M}{k}2N_t(N_t-1)$. However, this count includes the redundant terms that depend on only a single symbol.  A single-symbol term is generated from a product such as $t_{v,R}t_{v',R}$ only when a $k$-th order term from one component is multiplied by the zeroth-order (constant) term from the other. The number of $k$-th order terms in a single component's expansion is $\binom{M}{k}$ since it is equivalent to choose $k$ variables from $M$ candidates. Thus, for each of the four product types, $2\binom{M}{k}$ single-symbol terms are generated per pair of symbols. Summing over all pairs, the total number of redundant terms to be subtracted is $8\binom{M}{k} \cdot \frac{N_t(N_t-1)}{2} = \binom{M}{k}4N_t(N_t-1) = \binom{M}{k}2N_t(2N_t-2)$. Therefore, the number of terms involving exactly two transmitted symbols is this initial count minus the redundancy: $|\mathrm{A}| = \binom{2M}{k}2N_t(N_t-1) - \binom{M}{k}2N_t(2N_t-2)$.
Afterward, we count the terms in set $\mathrm{B}$. These terms, which depend on variables from only a single transmitted symbol, are generated by the second and third blocks of Eq. (38). The number of unique $k$-th order terms that can be formed from one of the two components (real or imaginary) of a single transmitted symbol is $\binom{M}{k}$. Since there are $N_t$ transmitted symbols, each with two components, the total count for this set is $|\mathrm{B}| = \binom{M}{k}2 N_t $.
Therefore, the total number of unique $k$-th order terms, $N_k^b$, is the sum $|\mathrm{A}| + |\mathrm{B}| = \binom{2M}{k}2N_{t}(N_{t}-1)-\binom{M}{k}2N_{t}(2N_{t}-3)$. 

In summary,  the number of $k$-th order terms in the objective function with binary variables $N_{k}^{\mathbf{b}}$ for the lower-order and high-order cases is written as follows: 
\begin{equation}
N_{k}^{\mathbf{b}}=\begin{cases}\binom{2M}{k}2N_{\mathrm{t}}(N_{t}-1)-\binom{M}{k}2N_{t}(2N_{t}-3) & \text{if } k \le M \\ \binom{2M}{k}2N_{\mathrm{t}}(N_{t}-1)& \text{if } k > M. \end{cases}
\end{equation}

Next, we derive the number of terms in the objective function with spin variables by analyzing the expansion of \eqref{eq:mimo_objfun_expansion}. The real and imaginary parts of a transmitted symbol $t'_v(s)$ are polynomials composed of $M$ distinct monomials, with orders ranging from $1$ to $M$. Let us denote the set of monomials in $t'_{v,R}$ as $\{\mu_{v,R,p}\}_{p=1}^M$ and in $t'_{v,I}$ as $\{\mu_{v,I,p}\}_{p=1}^M$, where the order of the monomial $\mu_{\cdot, \cdot, p}$ is $p$. The spin variables used for different symbols or for the real and imaginary parts of the same symbol are all distinct.

The first source of terms arises from the fourth block of the expansion, which contains products of polynomials from two different symbols, for example the product of $t'_{v,R}(s)$ and $t'_{v',R}(s)$. Since their variables are disjoint, the product of a $p$-th order monomial and a $q$-th order monomial results in a new monomial of order $p+q$. To count the generated $k$-th order terms, we find pairs $(p, q)$ where $p+q=k$ and $1 \le p, q \le M$. For low orders where $1 \le k \le M$, there are $k-1$ such pairs. For high orders where $M < k \le 2M$, there are $2M-k+1$ such pairs. There are four such product combinations for each of the $\frac{N_{\mathrm{t}}(N_{\mathrm{t}}-1)}{2}$ symbol pairs. Thus, this block contributes $4(k-1)\frac{N_{\mathrm{t}}(N_{\mathrm{t}}-1)}{2} = (k-1)2N_{\mathrm{t}}(N_{\mathrm{t}}-1)$ terms for low orders and
$4(2M-k+1)\frac{N_{\mathrm{t}}(N_{\mathrm{t}}-1)}{2}=(2M-k+1) 2N_{\mathrm{t}}(N_{\mathrm{t}}-1) $ for high orders.

In addition, we must account for terms that depend on the variables of a single symbol, which arise from the second and third blocks of the expansion. The third block contains squared polynomials for each symbol's real and imaginary parts, such as $(t'_{v,R})^2$ and $(t'_{v,I})^2$. Let us consider one such polynomial, for instance $t'_{v,R}$. When it is squared, it generates two types of terms: squares of each monomial, like $\mu_{v,R,p}^2$, and products between two distinct monomials, such as $\mu_{v,R,p}$ and $\mu_{v,R,q}$ where $p \neq q$. The same structure applies to the imaginary part. Due to the property $s_i^2=1$, the product of two distinct monomials with orders $p<q$ simplifies to a new monomial of order $q-p$. For a target order $k$, the number of pairs $(p,q)$ satisfying $q-p=k$ corresponds to the number of newly generated $k$-th order monomials from these products of distinct monomials, which is $M-k$.

Meanwhile, the second block of the expansion contributes the original linear polynomials, $t'_{v,R}$ and $t'_{v,I}$. Each of these polynomials itself contains exactly one monomial of order $k$. Therefore, for each single-symbol polynomial source, both real and imaginary, the total number of unique $k$-th order monomials is the sum of those newly generated from the products of distinct monomials, which is $M-k$, and the one original monomial from the linear term, resulting in a total of $M-k+1$ terms. Since there are $2N_{\mathrm{t}}$ such sources of terms in total, one for the real and one for the imaginary part of each of the $N_{\mathrm{t}}$ symbols, they contribute an additional $(M-k+1)2N_{\mathrm{t}}$ terms when $1 \le k \le M$.

By combining these contributions, the total number of unique $k$-th order terms in the objective function with spin variables, denoted by $N_k^{\mathbf{s}}$, is given by:
\begin{align}
N_k^{\mathbf{s}} =
\begin{cases}
    (k-1) 2N_{\mathrm{t}}(N_{\mathrm{t}}-1) + (M-k + 1)2N_{\mathrm{t}} & \text{if } k \le M, \\
    (2M-k+1) 2N_{\mathrm{t}}(N_{\mathrm{t}}-1) & \text{if } k > M.
\end{cases}
\end{align}

\footnotesize{
\bibliographystyle{IEEEtran}
\bibliography{main}

\begin{thebibliography}{10}
\providecommand{\url}[1]{#1}
\csname url@samestyle\endcsname
\providecommand{\newblock}{\relax}
\providecommand{\bibinfo}[2]{#2}
\providecommand{\BIBentrySTDinterwordspacing}{\spaceskip=0pt\relax}
\providecommand{\BIBentryALTinterwordstretchfactor}{4}
\providecommand{\BIBentryALTinterwordspacing}{\spaceskip=\fontdimen2\font plus
\BIBentryALTinterwordstretchfactor\fontdimen3\font minus \fontdimen4\font\relax}
\providecommand{\BIBforeignlanguage}[2]{{%
\expandafter\ifx\csname l@#1\endcsname\relax
\typeout{** WARNING: IEEEtran.bst: No hyphenation pattern has been}%
\typeout{** loaded for the language `#1'. Using the pattern for}%
\typeout{** the default language instead.}%
\else
\language=\csname l@#1\endcsname
\fi
#2}}
\providecommand{\BIBdecl}{\relax}
\BIBdecl

\bibitem{egger2020quantum}
D.~J. Egger, C.~Gambella, J.~Marecek, S.~McFaddin, M.~Mevissen, R.~Raymond, A.~Simonetto, S.~Woerner, and E.~Yndurain, ``Quantum computing for finance: State-of-the-art and future prospects,'' \emph{IEEE Transactions on Quantum Engineering}, vol.~1, pp. 1--24, 2020.

\bibitem{iasemidis2001quadratic}
L.~Iasemidis, P.~Pardalos, J.~Sackellares, and D.-S. Shiau, ``Quadratic binary programming and dynamical system approach to determine the predictability of epileptic seizures,'' \emph{Journal of Combinatorial Optimization}, vol.~5, no.~1, pp. 9--26, Mar. 2001.

\bibitem{sano2023qubit}
Y.~Sano, M.~Norimoto, and N.~Ishikawa, ``Qubit reduction and quantum speedup for wireless channel assignment problem,'' \emph{IEEE Transactions on Quantum Engineering}, vol.~4, pp. 1--12, 2023.

\bibitem{ibm2022user}
IBM, ``User's {{Manual}} for {{CPLEX}} 22.1.1,'' https://www.ibm.com/docs/en/icos/22.1.1, Dec. 2022.

\bibitem{gurobioptimizationllc2023gurobi}
G.~O. LLC, ``Gurobi {{Optimizer Reference Manual}},'' https://www.gurobi.com/, 2023.

\bibitem{luo2010semidefinite}
Z.-q. Luo, W.-k. Ma, A.~So, Y.~Ye, and S.~Zhang, ``Semidefinite relaxation of quadratic optimization problems,'' \emph{IEEE Signal Processing Magazine}, vol.~27, no.~3, pp. 20--34, May 2010.

\bibitem{ning2019optimization}
C.~Ning and F.~You, ``Optimization under uncertainty in the era of big data and deep learning: {{When}} machine learning meets mathematical programming,'' \emph{Computers \& Chemical Engineering}, vol. 125, pp. 434--448, Jun. 2019.

\bibitem{kadowaki1998quantum}
T.~Kadowaki and H.~Nishimori, ``Quantum annealing in the transverse ising model,'' \emph{Physical Review E}, vol.~58, no.~5, pp. 5355--5363, Nov. 1998.

\bibitem{farhi2014quantum}
E.~Farhi, J.~Goldstone, and S.~Gutmann, ``A quantum approximate optimization algorithm,'' \emph{arXiv:1411.4028}, Nov. 2014.

\bibitem{campbell2021qaoa}
C.~Campbell and E.~Dahl, ``{{QAOA}} of the highest order,'' \emph{arXiv:2111.12754 [quant-ph]}, Dec. 2021.

\bibitem{johnson2011quantum}
\BIBentryALTinterwordspacing
M.~W. Johnson, M.~H.~S. Amin, S.~Gildert, T.~Lanting, F.~Hamze, N.~Dickson, R.~Harris, A.~J. Berkley, J.~Johansson, P.~Bunyk, E.~M. Chapple, C.~Enderud, J.~P. Hilton, K.~Karimi, and E.~Ladiz, ``{Quantum annealing with manufactured spins},'' \emph{Nature}, vol. 473, no. 7346, pp. 194--198, May 2011. [Online]. Available: \url{https://ideas.repec.org/a/nat/nature/v473y2011i7346d10.1038_nature10012.html}
\BIBentrySTDinterwordspacing

\bibitem{andrew2019validating}
\BIBentryALTinterwordspacing
A.~W. Cross, L.~S. Bishop, S.~Sheldon, P.~D. Nation, and J.~M. Gambetta, ``Validating quantum computers using randomized model circuits,'' \emph{Physical Review A}, vol. 100, p. 032328, Sep 2019. [Online]. Available: \url{https://link.aps.org/doi/10.1103/PhysRevA.100.032328}
\BIBentrySTDinterwordspacing

\bibitem{stilckfranca2021limitations}
D.~Stilck~Fran{\c c}a and R.~{Garc{\'i}a-Patr{\'o}n}, ``Limitations of optimization algorithms on noisy quantum devices,'' \emph{Nature Physics}, vol.~17, no.~11, pp. 1221--1227, Nov. 2021.

\bibitem{gilliam2021grover}
A.~Gilliam, S.~Woerner, and C.~Gonciulea, ``Grover adaptive search for constrained polynomial binary optimization,'' \emph{Quantum}, vol.~5, p. 428, Apr. 2021.

\bibitem{gilliam2020optimizing}
A.~Gilliam, M.~Pistoia, and C.~Gonciulea, ``Optimizing quantum search with a binomial version of grover's algorithm,'' \emph{arXiv:2007.10894}, Jul. 2020.

\bibitem{grover1996fast}
L.~K. Grover, ``A fast quantum mechanical algorithm for database search,'' in \emph{Proceedings of the Twenty-Eighth Annual {{ACM}} Symposium on {{Theory}} of {{Computing}}}, New York, NY, USA, Jul. 1996, pp. 212--219.

\bibitem{norimoto2023quantum}
M.~Norimoto, R.~Mori, and N.~Ishikawa, ``Quantum algorithm for higher-order unconstrained binary optimization and {{MIMO}} maximum likelihood detection,'' \emph{IEEE Transactions on Communications}, vol.~71, no.~4, pp. 1926--1939, Apr. 2023.

\bibitem{gilliam2021foundational}
A.~Gilliam, C.~Venci, S.~Muralidharan, V.~Dorum, E.~May, R.~Narasimhan, and C.~Gonciulea, ``Foundational patterns for efficient quantum computing,'' \emph{arXiv:1907.11513}, Jan. 2021.

\bibitem{nagy2024fixedpoint}
{\'A}.~Nagy, J.~Park, C.~Zhang, A.~Acharya, and A.~Khan, ``Fixed-point {{Grover}} adaptive search for binary optimization problems,'' \emph{arXiv:2311.05592}, May 2024.

\bibitem{lucas2014ising}
A.~Lucas, ``Ising formulations of many {{NP}} problems,'' \emph{Frontiers in Physics}, vol.~2, 2014.

\bibitem{fuchs2021efficient}
F.~G. Fuchs, H.~{\O}. Kolden, N.~H. Aase, and G.~Sartor, ``Efficient encoding of the weighted max $k$-cut on a quantum computer using {{QAOA}},'' \emph{SN Computer Science}, vol.~2, no.~2, p.~89, Feb. 2021.

\bibitem{glos2022spaceefficient}
A.~Glos, A.~Krawiec, and Z.~Zimbor{\'a}s, ``Space-efficient binary optimization for variational quantum computing,'' \emph{npj Quantum Information}, vol.~8, no.~1, pp. 1--8, Apr. 2022.

\bibitem{sano2024accelerating}
Y.~Sano, K.~Mitarai, N.~Yamamoto, and N.~Ishikawa, ``Accelerating grover adaptive search: Qubit and gate count reduction strategies with higher order formulations,'' \emph{IEEE Transactions on Quantum Engineering}, vol.~5, pp. 1--12, 2024.

\bibitem{giuffrida2022engineering}
L.~Giuffrida, D.~Volpe, G.~A. Cirillo, M.~Zamboni, and G.~Turvani, ``Engineering grover adaptive search: Exploring the degrees of freedom for efficient {{QUBO}} solving,'' \emph{IEEE Journal on Emerging and Selected Topics in Circuits and Systems}, vol.~12, no.~3, pp. 614--623, Sep. 2022.

\bibitem{ominato2024grover}
H.~Ominato, T.~Ohyama, and K.~Yamaguchi, ``Grover adaptive search with fewer queries,'' \emph{IEEE Access}, vol.~12, pp. 74\,619--74\,632, 2024.

\bibitem{bennett1997strengths}
C.~H. Bennett, E.~Bernstein, G.~Brassard, and U.~Vazirani, ``Strengths and weaknesses of quantum computing,'' \emph{SIAM Journal on Computing}, vol.~26, no.~5, pp. 1510--1523, Oct. 1997.

\bibitem{Jones1997Computability}
N.~D. Jones, \emph{Computability and Complexity: From a Programming Perspective}.\hskip 1em plus 0.5em minus 0.4em\relax MIT Press, 1997, vol.~21.

\bibitem{zhu2022realizable}
J.~Zhu, Y.~Gao, H.~Wang, T.~Li, and H.~Wu, ``A realizable {GAS}-based quantum algorithm for traveling salesman problem,'' \emph{arXiv.2212.02735}, 12 2022.

\bibitem{Krol2024QISS}
A.~M. Krol, M.~Erdmann, R.~Mishra, P.~Singkanipa, E.~Munro, M.~Ziolkowski, A.~Luckow, and Z.~Al-Ars, ``Qiss: Quantum industrial shift scheduling algorithm,'' \emph{arXiv:2401.07763}, 2024.

\bibitem{norimoto2024quantum}
M.~Norimoto, T.~Mikuriya, and N.~Ishikawa, ``Quantum speedup for multiuser detection with optimized parameters in grover adaptive search,'' \emph{IEEE Access}, vol.~12, pp. 83\,810--83\,821, 2024.

\bibitem{yukiyoshi2022quantum}
K.~Yukiyoshi and N.~Ishikawa, ``Quantum search algorithm for binary constant weight codes,'' \emph{arXiv:2211.04637}, Nov. 2022.

\bibitem{yukiyoshi2024quantum}
K.~Yukiyoshi, T.~Mikuriya, H.~Rou, G.~de~Abreu, and N.~Ishikawa, ``Quantum speedup of the dispersion and codebook design problems,'' \emph{IEEE Transactions on Quantum Engineering}, vol.~5, no.~01, pp. 1--16, jan 2024.

\bibitem{3gpp2018ts}
{3GPP}, ``{{TS}} 138 211 - v15.2.0 - {{5G}}; {{NR}}; physical channels and modulation ({{3GPP TS}} 38.211 version 15.2.0 release 15),'' 2018.

\bibitem{Silva2023LinearDecomposition}
J.~D.~S. Silva, T.~M.~D. Azevedo, I.~F. Araujo, and A.~J. da~Silva, ``Linear decomposition of approximate multi-controlled single qubit gates,'' \emph{arXiv:2310.14974}, 2023.

\bibitem{McKay2017Efficient}
D.~C. McKay, C.~J. Wood, S.~Sheldon, J.~M. Chow, and J.~M. Gambetta, ``{Efficient Z-gates for Quantum Computing},'' \emph{Physical Review A}, vol.~96, p. 022330, 2017.

\bibitem{silva2022linear}
\BIBentryALTinterwordspacing
A.~J. da~Silva and D.~K. Park, ``Linear-depth quantum circuits for multiqubit controlled gates,'' \emph{Physical Review A}, vol. 106, p. 042602, Oct 2022. [Online]. Available: \url{https://link.aps.org/doi/10.1103/PhysRevA.106.042602}
\BIBentrySTDinterwordspacing

\bibitem{verdu1989computationa}
\BIBentryALTinterwordspacing
S.~Verd\'{u}, ``Computational complexity of optimum multiuser detection,'' \emph{Algorithmica}, vol.~4, no. 1–4, p. 303–312, mar 1989. [Online]. Available: \url{https://doi.org/10.1007/BF01553893}
\BIBentrySTDinterwordspacing

\bibitem{Abrasao2010MIMO}
T.~Abrão, L.~D. de~Oliveira, F.~Ciriaco \emph{et~al.}, ``{S/MIMO MC-CDMA} heuristic multiuser detectors based on single-objective optimization,'' \emph{Wireless Personal Communications}, vol.~53, p. 529–553, 2010.

\bibitem{berlekamp1978inherent}
E.~Berlekamp, R.~McEliece, and H.~van Tilborg, ``On the inherent intractability of certain coding problems (corresp.),'' \emph{IEEE Transactions on Information Theory}, vol.~24, no.~3, pp. 384--386, 1978.

\bibitem{mceliece1978public}
R.~J. McEliece, ``A public-key cryptosystem based on algebraic coding theory,'' \emph{DSN Progress Report}, vol. 42--44, pp. 114--116, 1978.

\bibitem{bartschi2019deterministic}
A.~B{\"a}rtschi and S.~Eidenbenz, ``Deterministic preparation of {{Dicke}} states,'' in \emph{Fundamentals of {{Computation Theory}}}, 2019, vol. 11651, pp. 126--139.

\bibitem{botsinis2014fixedcomplexity}
P.~Botsinis, S.~X. Ng, and L.~Hanzo, ``Fixed-complexity quantum-assisted multi-user detection for {{CDMA}} and {{SDMA}},'' \emph{IEEE Transactions on Communications}, vol.~62, no.~3, pp. 990--1000, Mar. 2014.

\end{thebibliography}
}

\begin{IEEEbiography}{Shintaro~Fujiwara}
(Graduate Student Member, IEEE)  received the B.E. degree in engineering science in 2024 from Yokohama National University, Kanagawa, Japan, where he is currently working toward the M.E. degree in
electrical and computer engineering.
His research interests include quantum algorithms, coding theory, and wireless communications
\end{IEEEbiography}

\begin{IEEEbiography}{Naoki~Ishikawa} (S'13--M'17--SM'22) is an Associate Professor with the Faculty of Engineering, Yokohama National University, Kanagawa, Japan. He received the B.E., M.E., and Ph.D. degrees from Tokyo University of Agriculture and Technology, Tokyo, Japan, in 2014, 2015, and 2017, respectively. In 2015, he was an academic visitor with the School of Electronics and Computer Science, University of Southampton, U.K. From 2016 to 2017, he was a research fellow of the Japan Society for the Promotion of Science. From 2017 to 2020, he was an Assistant Professor in the Graduate School of Information Sciences, Hiroshima City University, Japan. From 2025 to 2026, he was a research scholar with the Jacobs School of Engineering, University of California, San Diego, U.S. He was certified as an Exemplary Reviewer of \textsc{IEEE Transactions on Communications} in 2017 and 2021. His research interests include quantum algorithms and wireless communications.
\end{IEEEbiography}

\EOD

\end{document}